\documentclass[12pt]{article}
\usepackage{epsf} 
\usepackage{amssymb}
\usepackage{color}
\usepackage[dvips]{graphics} 
\author{Alex Roxin, Hermann Riecke\\
{\small Department of Engineering Sciences and Applied Mathematics,}\\
 {\small Northwestern University, 2145 Sheridan Rd}\\
 {\small Evanston, IL, 60208, USA}\\
 {\small tel: (847)-491-5396}\\
 {\small fax: (847)-491-2178}\\
 {\small email: h-riecke@nwu.edu, a-roxin@nwu.edu}\small}
	  
\title{Destabilization and Localization of Traveling Waves by an Advected Field}
\begin{document}
\bibliographystyle{unsrt}

\maketitle
\begin{abstract}
{We study a model of small-amplitude traveling waves arising in a supercritical Hopf-bifurcation, that are coupled to
a slowly varying, real field.  The field is advected by the waves and, in turn, affects their stability via a
coupling to the growth rate.  In the absence of dispersion we identify two distinct shortwave instabilities.
One instability induces a phase slip of the waves and a corresponding reduction of the winding number, while
the other leads to a modulated wave structure.  The bifurcation to modulated waves can be either forward
or backward, in the latter case permitting the existence of localized, traveling pulses which are bistable 
with the  basic, conductive state.} 
\end{abstract}

\section{Introduction}

Various systems have been shown to exhibit bifurcations to spatially periodic patterns in the presence of an 
additional, weakly damped mode.  If that mode corresponds to a conserved quantity it constitutes a true
zero mode as in, e.g., translation of the fluid interface in two-layer Couette flow \cite{ChBa99} or a shift in
the displacement velocity of seismic waves in a viscoelastic medium \cite{Ma92a}.  Examples of weakly damped
modes include the large-scale concentration field in binary-fluid convection \cite{Ri92,Ri92a} or the real, 
slow mode in the  4-species Oregonator model of the BZ reaction \cite{IpSo00}.  The evolution of instabilities 
in these and similar systems will be coupled to the dynamics of the weakly-damped mode.
Within the context of a weakly nonlinear approach near onset of the pattern-forming instability, the
 Ginzburg-Landau model is altered through the coupling to an evolution equation for the additional mode.
The precise form of the coupling depends on the physics and symmetries of the system. 

 Steady instabilities in conserved
systems with and without reflection symmetry have been considered.
With reflection symmetry the instability, setting in at finite wavenumber, was found by Mathews and 
Cox \cite{MaCo00} to be amplitude-driven, leading to a supercritical modulation of the underlying pattern.  In the absence of 
reflection symmetry, Malomed \cite{Ma92a} identified modulated patterns by deriving longwave equations for the 
phase of the underlying
pattern coupled to the zero-mode.  A particle-in-a-potential model in a moving frame was then derived for the local wavenumber, 
which admits both traveling and steady modulations.  Ipsen and S\o rensen \cite{IpSo00} investigated the effect of a real, 
slow mode in reaction-
diffusion systems near a supercritical Hopf bifurcation.  They found the slow mode leads to new finite-wavenumber instabilities
which alter traditional Eckhaus and Benjamin-Feir stability criteria for periodic waves.  Barthelet and Charru \cite{BaCh98}
studied instabilities
of interfacial waves in two-layer Couette-Poiseuille flow with 
Galilean invariance and no reflection symmetry, where the zero-mode is a shift of the fluid interface.  The 
corresponding model of 
waves coupled to the zero-mode, derived and worked out 
by Renardy and Renardy \cite{ReRe93}, results in multiple, separated stability
regions for periodic waves in contrast to the Eckhaus stable band exhibited by a single Ginzburg-Landau equation.  The critically
damped mode corresponding to a two-dimensional mean-flow in stress-free convection 
becomes relevant for not too large values of the Prandtl
number.  Bernoff \cite{Be94} derived a Ginzburg-Landau equation coupled to the mean-flow mode from the Bousinesq equations and 
calculated the skew-varicose and oscillatory skew-varicose stability boundaries.  This calculation brought into agreement 
previous work by both Siggia and Zippelius \cite{SiZi81} and Busse and Bolton \cite{BuBo84} who investigated the effect 
of mean-flow modes with non-zero vertical vorticity on the stability of rolls in Rayleigh-Benard convection.  

In binary-fluid convection, Riecke \cite{Ri92}
derived an amplitude
equation for traveling waves arising in a Hopf bifurcation coupled to a critically damped mode related to large-scale 
modulations of the 
concentration field in the limit of small Lewis number.  Instabilities of periodic waves within these equations were found to lead to
localized traveling pulses which exhibited the anomalous slow drift observed in experiment \cite{Ko91}, and in numerical
simulations \cite{BaLu91}.
  Riecke and Granzow presented a similar model on phenomenological grounds to describe the dynamics of traveling waves in 
electroconvection in nematic liquid crystals \cite{RiGr98}.  This model for oblique (zig and zag) waves coupled to a slow mode,
 possibly corresponding to a charge-carrier mode, exhibits instabilities to localized, worm-like structures as seen in experiment 
\cite{DeAh96a}.  A striking feature of the worms is that although the extended waves bifurcate supercritically from the basic, 
conductive state, the worms themselves are bistable with the latter.

In the present paper, we consider traveling waves arising in a supercritical Hopf-bifurcation coupled to a real, slowly
 varying field.  The field is advected by the waves and, in turn, can affect the dynamics of the waves through coupling
to their growth rate.  We briefly introduce the model and  present a cursory linear analysis in section 2 which reveals 
distinct phase and amplitude instabilities of the waves.  In sections 3 and 4 we are concerned with characterizing the
linear and nonlinear behaviors of the phase and amplitude instabilities, respectively.  In both cases we derive an envelope
equation for the instability near its threshold and compare the results to full numerical simulations of the original equations.  The
envelope equations, in agreement with the numerics, indicate the phase-instability leads to a backward Hopf bifurcation, while
the amplitude-driven instability leads to modulated waves, arising super- or subcritically.  In the latter case, the subcritical branch
can be bistable with the basic, conductive state and localized wave pulses arise.

\section{The Extended Ginzburg-Landau Equations}

We consider a model of a traveling wave with complex amplitude A and group velocity s coupled
 to a real, weakly damped
mode C.  The general form of the equations to orders $\epsilon^3, \epsilon^4$ respectively, where $\epsilon$ is a
 measure of the distance from threshold of the bifurcation to traveling waves ($a_{1}=O(\epsilon^{2}$)), is dictated by symmetry,
\begin{eqnarray}
\partial_{t}A+s\partial_{x}A&=&d_{1}\partial_{x}^{2}A+(a_{1}+a_{2}C)A-b|A|^{2}A, \label{A}\\
\partial_{t}C&=&d_{2}\partial_{x}^{2}C-a_{3}C+a_{4}C^{2}+(h_{1}+h_{2}C)|A|^{2}+h_{3}|A|^{4}+h_{4}\partial_{x}|A|^{2}+
h_{5}\partial_{x}^{2}|A|^{2}\nonumber\\
&&+h_{6}i(\bar{A}\partial_{x}A-A\partial_{x}\bar{A})+h_{7}i\partial_{x}(\bar{A}\partial_{x}A-A\partial_{x}\bar{A})
+h_{8}\partial_{x}A\partial_{x}\bar{A}. \label{C}
\end{eqnarray}
The coefficients $d_{1}, a_{1}, a_{2}, b$ in (1) are, in general, complex.  The coefficients can be calculated 
 for a particular system through a perturbative, normal-mode  expansion of the relevant variables with slowly varying amplitudes,
\begin{eqnarray}
\mathbf{\Phi}
=\epsilon (\mathbf{u}A(x,t)e^{i(q\tilde{x}+\omega \tilde{t})}+c.c.)+\epsilon^{2}(\mathbf{v}C(x,t)+...)+h.o.t. 
\end{eqnarray}
We consider a field $C$, that relaxes to a unique state in the absence of forcing by the wave and thus consider the contribution
of the term proportional to $a_{4}$, which could introduce a second branch through a transcritical bifurcation, to be of higher order.  
We furthermore focus on the affect of advection by the traveling waves on C by retaining only the gradient coupling in (\ref{C})
and eliminating the dispersion in (\ref{A}).  In the case of thermal binary fluid convection it turns out that the term proportional
to $h_{1}$, which is formally of lower order than the gradient term, only contributes at higher order \cite{Ri96}.  A gradient coupling  of 
counterpropagating, traveling waves to a slowly varying field as a model for electroconvection in nematic liquid crystals has 
reproduced localized wave trains or 'worms' seen in experiment \cite{RiGr98}.  In systems with a true zero
mode (conserved quantity) certain terms like those involving $h_{1}, h_{2}, h_{3}$ and $h_{5}$ are not allowed and the gradient coupling is the 
relevant forcing \cite{ChBa99}, \cite{MaCo00}.  

With the abovementioned simplifications, the system can be rewritten as,
\begin{eqnarray}
\partial_{t}A+s\partial_{x}A&=&\partial_{x}^{2}A+(a+C)A-|A|^{2}A,  \label{EGLE:A}\\
\partial_{t}C&=&\delta\partial_{x}^{2}C-\alpha C+h\partial_{x}|A|^{2}. \label{EGLE:C}
\end{eqnarray}

Equations (\ref{EGLE:A}, \ref{EGLE:C}) describe the dynamics of a traveling wave without dispersion, which advects a 
real, slowly varying field C. The strength and sign of advection is given by the coefficient \textit{h}.  No homogeneous  scaling is possible
in these equations due to the presence of gradient terms which contribute to lower order in the limit of large-scale modulations.
  The scalings $\partial_{x} = O(\epsilon), \partial_{t} = O(\epsilon^{2})$ require $s=O(\epsilon), h=O(\epsilon)$ for strict 
validity of (\ref{EGLE:A},\ref{EGLE:C}) indicating small group velocity and weak coupling.
 Modulations on a longer length scale ($\partial_{x} = O(\epsilon^{2})$) allow for 
O(1) values of \textit{s} and \textit{h} although the diffusive terms now contribute at higher order.  The system thus becomes hyperbolic
in this limit.  Counterpropagating waves near onset have been studied in the hyperbolic limit \cite{MaVe98}, and can exhibit both
sub- and supercritical secondary bifurcations as well as more complex dynamics.  In this paper we choose not to reduce the equations
(\ref{EGLE:A},\ref{EGLE:C}) to a simpler form in a distinguished limit by fixing the scale.  Instead we retain all terms as O(1)
quantities and consider the equations as a phenomenological model of traveling waves coupled to a critically damped mode.

Our goal is to identify instabilities of waves in (\ref{EGLE:A},\ref{EGLE:C})
 and characterize their nonlinear evolution.  We note that in the absence of coupling (\textit{h}=0) or in the limit of rapid
decay of the C-field ($\alpha\to\infty$) (\ref{EGLE:A},\ref{EGLE:C}) reduce to a single Ginzburg-Landau equation 
which exhibits a long-wave phase
instability for waves with wavenumber $q^{2} > \frac{a}{3}$.  We shall refer to (\ref{EGLE:A},\ref{EGLE:C}) as the extended Ginzburg-Landau 
equations (EGLE).

The complex amplitude \textit{A} in (\ref{EGLE:A}) can be described by its real magnitude and phase $A=Re^{i\theta}$.
Thus, in a frame moving with the waves (\ref{EGLE:A},\ref{EGLE:C}) can be decomposed into the three following coupled equations,

\begin{eqnarray}
\partial_{t}R&=&\partial_{x}^{2}R+R[(a+C)-R^{2}-(\partial_{x}\theta)^{2}], \label{EGLE:R}\\
R^{2}\partial_{t}\theta&=&\partial_{x}[R^{2}\partial_{x}\theta], \label{EGLE:theta}\\
\partial_{t}C&=&\delta\partial_{x}^{2}C-\alpha C+s\partial_{x}C+2hR\partial_{x}R. \label{EGLE:C2}
\end{eqnarray}

To determine instabilities of the traveling waves we linearize about the wave solution
by considering an ansatz,

\begin{eqnarray}
R&=&\sqrt{a-q^{2}}+\epsilon\tilde{R}e^{ipx+\sigma t}, \label{lin:R}\\
\theta &=&qx+\epsilon\tilde{\theta} e^{ipx+\sigma t}, \label{lin:theta}\\
C&=&\epsilon\tilde{C} e^{ipx+\sigma t}. \label{lin:C}
\end{eqnarray}

We first consider instabilities in the longwave limit by expanding the growth rate $\sigma$ for small $p$.  The three
eigenvalues, corresponding to the real amplitude, phase and $C$-field respectively, are,

\begin{eqnarray}
\sigma_{amp}(p)&=&-2R^{2}-2\frac{ihR^{2}}{(2R^{2}-\alpha)}p\nonumber\\
&&-\frac{[4R^{6}\beta-2shR^{4}\alpha +(R^{2}+2q^{2})(2R^{2}-\alpha)^{3}]}{R^{2}(2R^{2}-\alpha)^{3}}p^{2}+O(p^{3}), \label{eig:R}\\
\sigma_{phase}(p)&=&-D_{E}p^{2}+2
\frac{ihq^{2}}{\alpha R^{2}}p^{3}-2\frac{q^{2}[\beta
+\frac{\alpha^{2}q^{2}}{R^{4}}]}{\alpha^{2}R^{2}}p^{4}+O(p^{5}), \label{eig:theta}\\
\sigma_{field}(p)&=&-\alpha +\frac{i(2R^{2}(s+h)-s\alpha)}{(2R^{2}-\alpha)}p\nonumber\\
&&-\frac{[-4R^{4}\beta+2shR^{2}\alpha +(2R^{2}-\alpha)^{3}\delta]}{(2R^{2}-\alpha)}p^{2}+O(p^{3}). \label{eig:C}
\end{eqnarray}

Here $R^{2}=\sqrt{a-q^{2}}$ is the amplitude of the plane wave and,

\begin{equation}
D_{E}=\frac{R^{2}-2q^{2}}{R^{2}}.\label{D}  
\end{equation}

The parameter $\beta =
h(s+h)$ is introduced for simplicity; it will prove to be useful in characterizing the type of instability.

We notice that to leading order, the eigenvalues corresponding to the amplitude (\ref{eig:R}) and 
advected field (\ref{eig:C}) are negative for finite values
of $R^{2}$ and $\alpha$.  The critical mode, corresponding to the phase, exhibits a longwave instability when the diffusion
coefficient $D_{E}$ changes sign; this occurs at the Eckhaus curve $q^{2}=\frac{a}{3}$.  However, the coupling of the advected
field now raises the possibility of the quartic order term balancing the quadratic term in (12) as $D_{E}\to 0$.  This would indicate
a small-, yet finite-wavenumber instability.  Indeed, a necessary condition for this phase instability to occur is
that $ \beta+\frac{\alpha^{2}q^{2}}{R^{4}}< 0$.  Figure 1(a) shows the eigenvalue of the phase mode which exhibits this finite-
wavenumber instability, i.e. sufficiently small wavenumbers are damped.  In Figure 1(c) we see the corresponding linear stability diagram. Here
plane waves become linearly unstable already before the Eckhaus curved is reached.  However, we note that near onset of the traveling
waves, i.e. as $a\to 0$ for finite $\alpha$, (\ref{EGLE:A}, \ref{EGLE:C}) can be rescaled, and $C$ adiabatically eliminated to
yield a single Ginzburg-Landau equation.  In this limit the contribution of the C-field is formally of higher order and we
recover the Eckhaus instability (cf. Fig. \ref{Fig:linstabint}(a), below).

\begin{figure}[t!]
\label{fig:lin}
\centerline{\resizebox{3.2in}{!}{\includegraphics{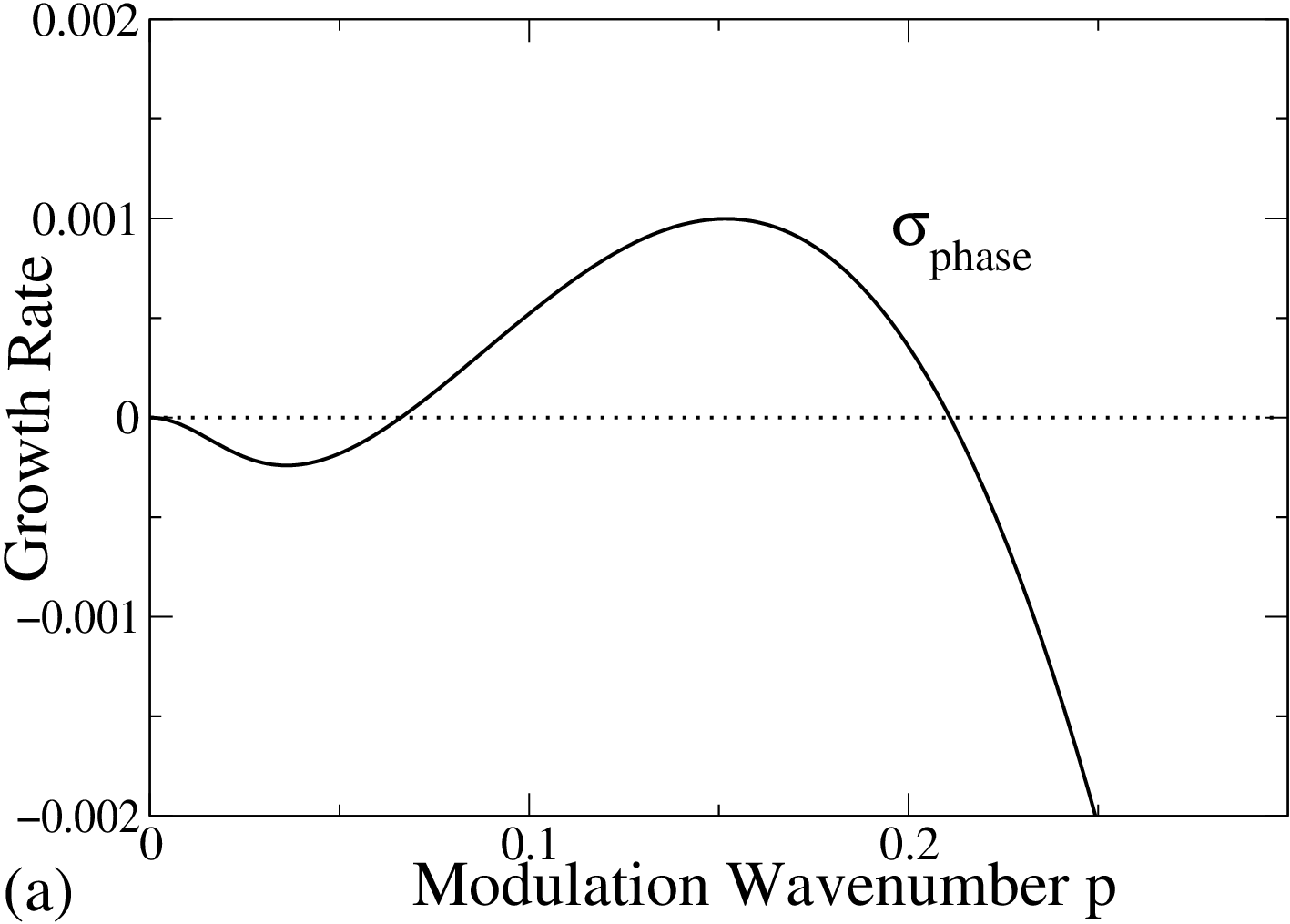}}\hspace{0.2in}\resizebox{3.2in}{!}{\includegraphics{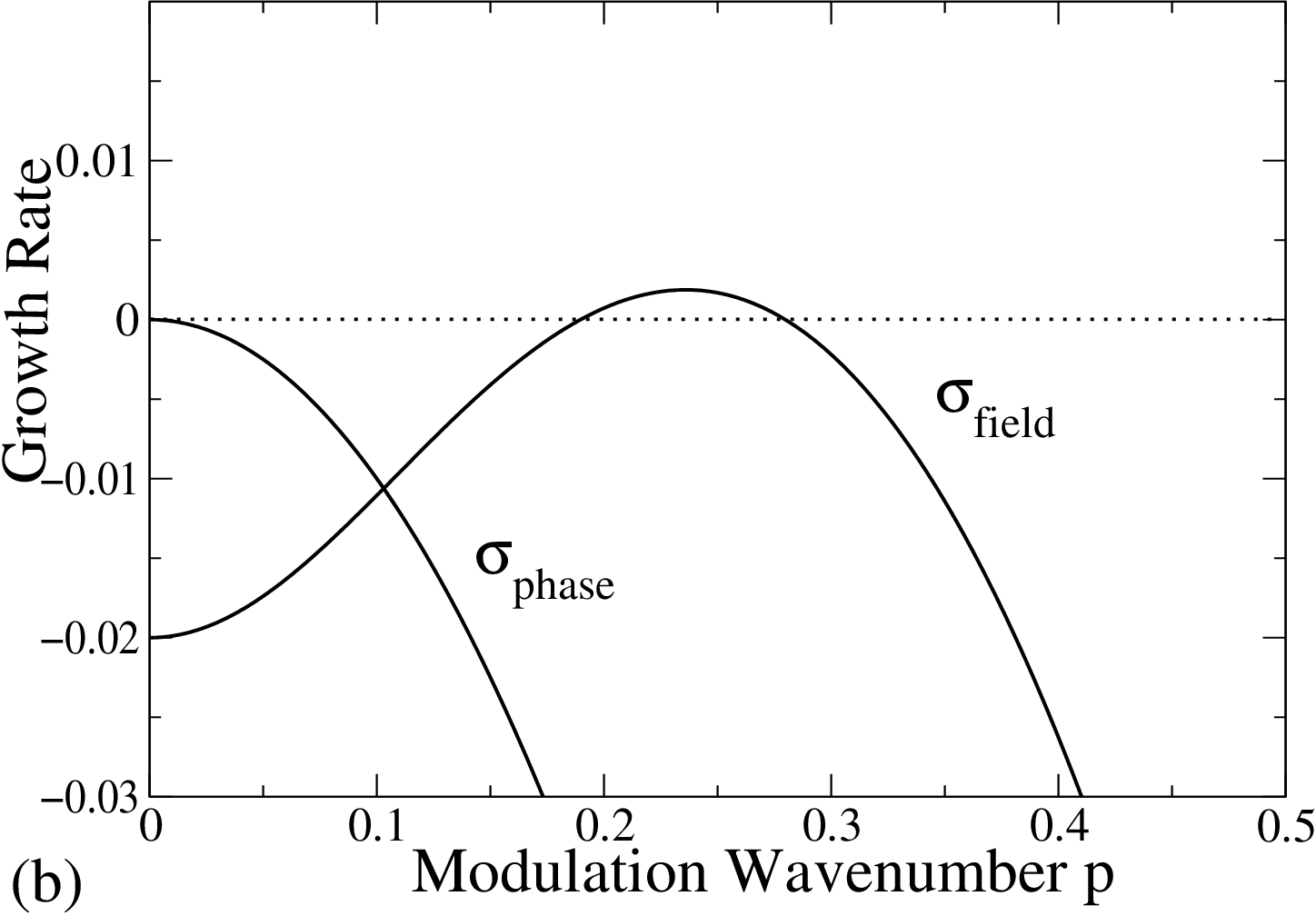}}}
\vspace{0.4in}\resizebox{3.2in}{!}{\includegraphics{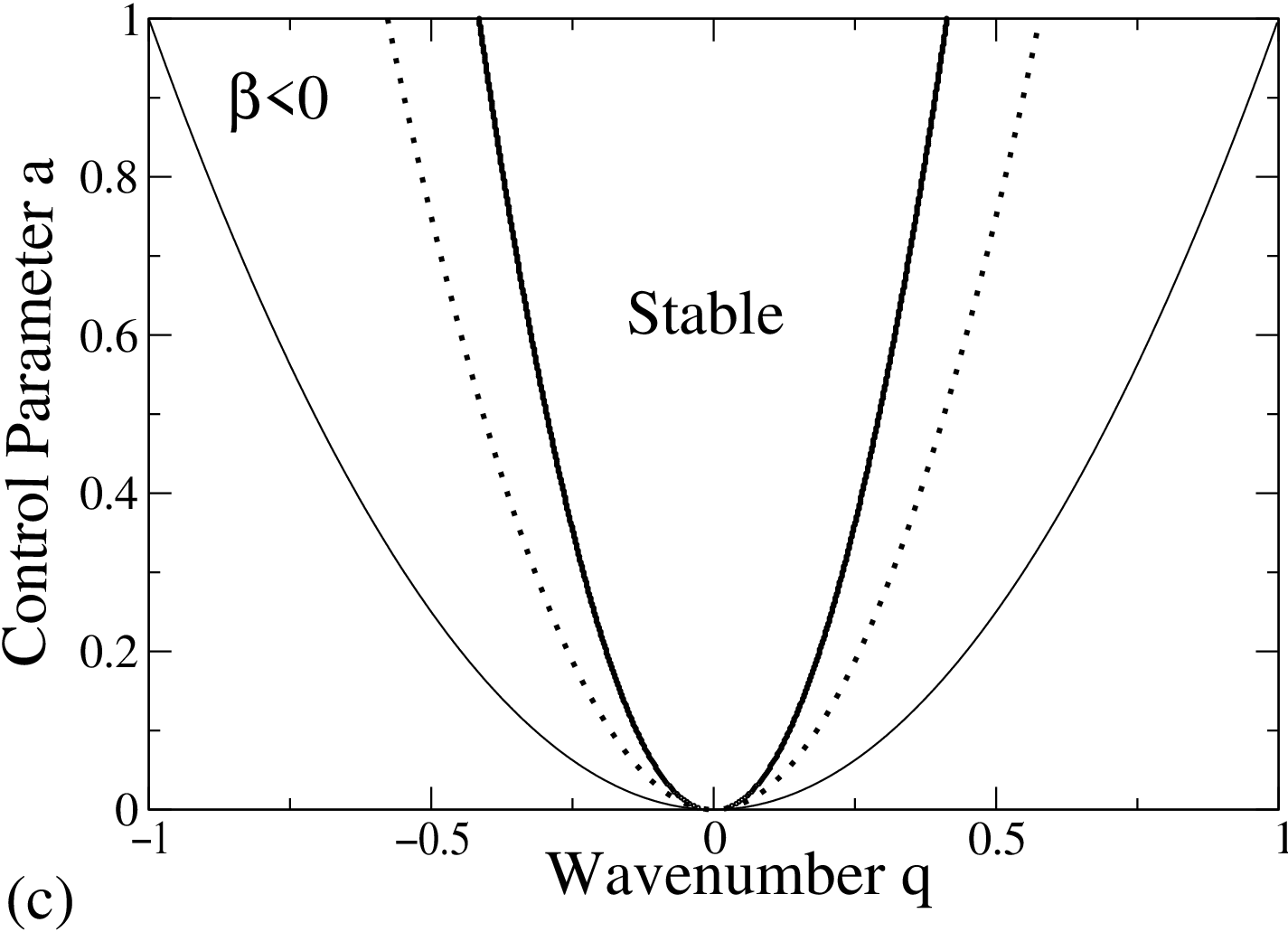}}\hspace{0.2in}\resizebox{3.2in}{!}{\includegraphics{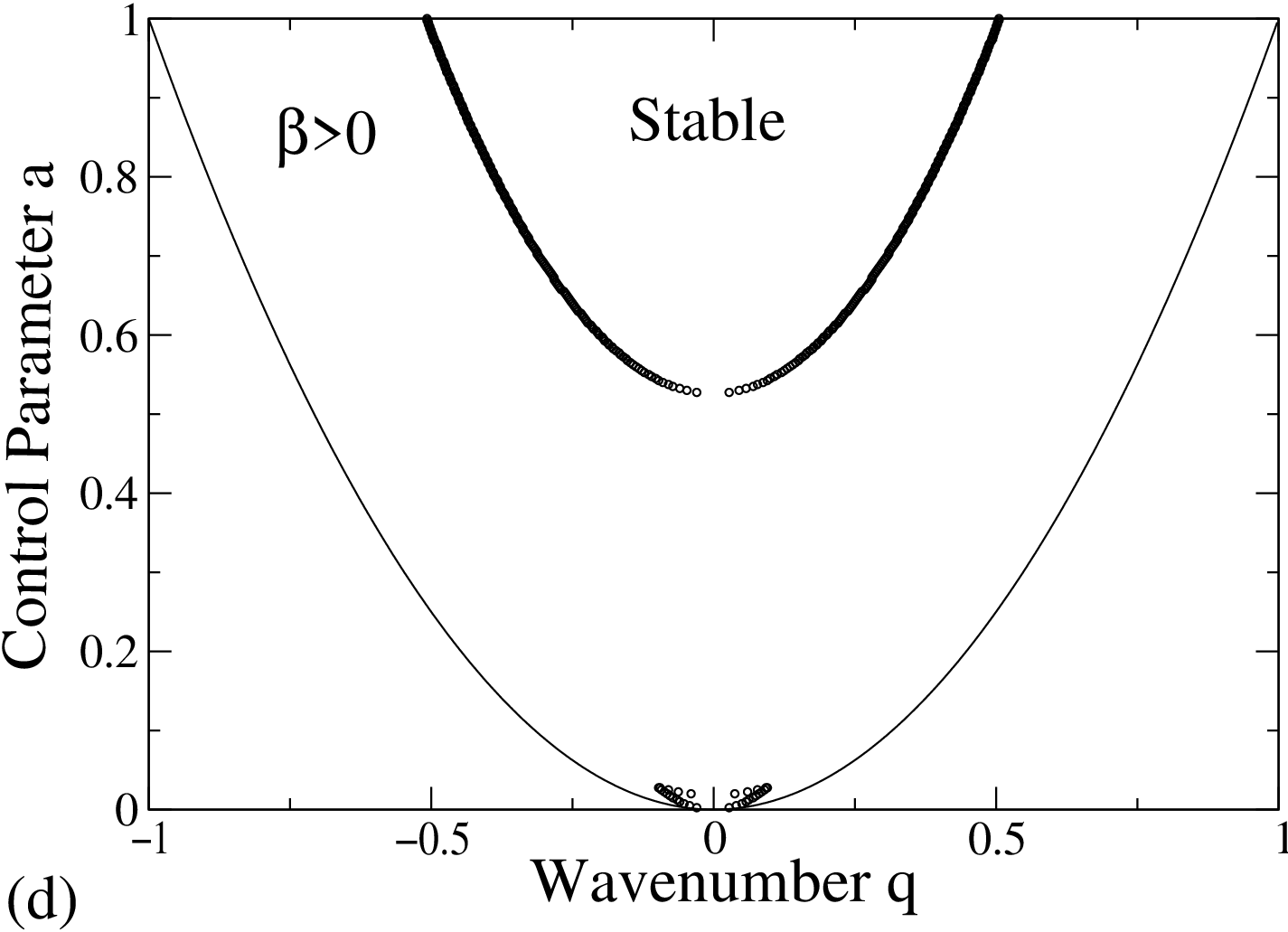}}
\caption{For all 4 figures h=1, $\alpha$=0.02, $\delta$=1.0  In (a) s=-1.5, q=0.425, a=1.0. This is a shortwave phase
instability.  The other two eigenvalues are large and negative.  In (b) s=1.0, q=0, a=0.5. This is a shortwave instability of the C-field.  The
phase-eigenvalue is marginal for p=0 due to translation symmetry while the amplitude-eigenvalue is large and negative. (c) is the linear 
stability diagram for the values of the coefficients given in (a).  The dotted line is the Eckhaus curve, given here for comparison.  (d)
is the linear stability diagram for the same coefficients as in (b).} \label{Fig:linstabgen}
\end{figure}

Next consider waves which are phase-stable in this regime ($D_{E}>O(p^{2})$ or $ \beta+\frac{\alpha^{2}q^{2}}{R^{4}}> 0$ or both).  In
the limit as $\alpha\to 0$, the eigenvalue corresponding to the C-field can become positive through a balance between
the leading order term and that at $O(p^{2})$.  Consider, for example, the eigenvalues at bandcenter (q=0), 

\begin{eqnarray}
\sigma_{amp}(p)&=&-2a-2\frac{iha}{(2a-\alpha)}p-\frac{[4a^{2}\beta-2sha\alpha +(2a-\alpha)
^{3}]}{(2a-\alpha)^{3}}p^{2}+O(p^{3}), \label{eig:Rbc}\\
\sigma_{phase}(p)&=&-p^{2}, \label{eig:thetabc}\\
\sigma_{field}(p)&=&-\alpha +\frac{i(2a(s+h)-s\alpha)}{(2a-\alpha)}p-\frac{[-4a^{2}\beta+2sha\alpha 
+(2a-\alpha)^{3}\delta]}{(2a-\alpha)}p^{2}+O(p^{3}). \label{eig:Cbc}
\end{eqnarray}

While for finite \textit{a} as $\alpha\to 0$, both the amplitude and phase eigenvalues are stable the 
field-eigenvalue can
become unstable if $-4\beta+8a\delta < 0$.  This condition can only be satisfied for $\beta >0$ and implies additionally that there
is an upper bound on the control parameter \textit{a} for fixed $\beta$ and $\delta$ above which this instability cannot 
exist ($a_{bound}=\frac{\beta}{2\delta}$).  Such a case is seen in Figure \ref{Fig:linstabgen}(b) where the eigenvalue corresponding to modulations of
the C-field indicates positive growth for finite-wavenumber perturbations.  The linear stability diagram Figure \ref{Fig:linstabgen}(d) confirms the 
existence of the linear instability over a finite range of the control parameter $a$, resulting in a small stability-island near onset.

 As $\beta$ passes from negative values through zero, the linear stability diagram given in Figure \ref{Fig:linstabgen}(c)
changes continuously to that shown in \ref{Fig:linstabgen}(d).  The intermediate regime is not captured analytically in the longwave expansions we
have used thus far and must be investigated numerically.  
Figure \ref{Fig:linstabint}(b) shows how the region of linear stability of the traveling waves changes as $\beta$ passes through zero.  The stability boundary,
outside of which the waves are phase-unstable to finite-wavenumber modulations for $\beta < 0$, develops a bottle-necked region which
pinches off as $\beta$ becomes more positive, resulting in two separated plane-wave-stable regions (cf. Fig.\ref{Fig:linstabgen}(d)).
  This change is continuous, and there are values of 
$\beta$ for which numerical simulations have revealed both phase- and amplitude-instabilities.  That is, amplitude-modulated waves can 
arise whose underlying phase undergoes slips until a phase-stable wavenumber is achieved.

\begin{figure}[t!]
\centerline{\resizebox{3.2in}{!}{\includegraphics{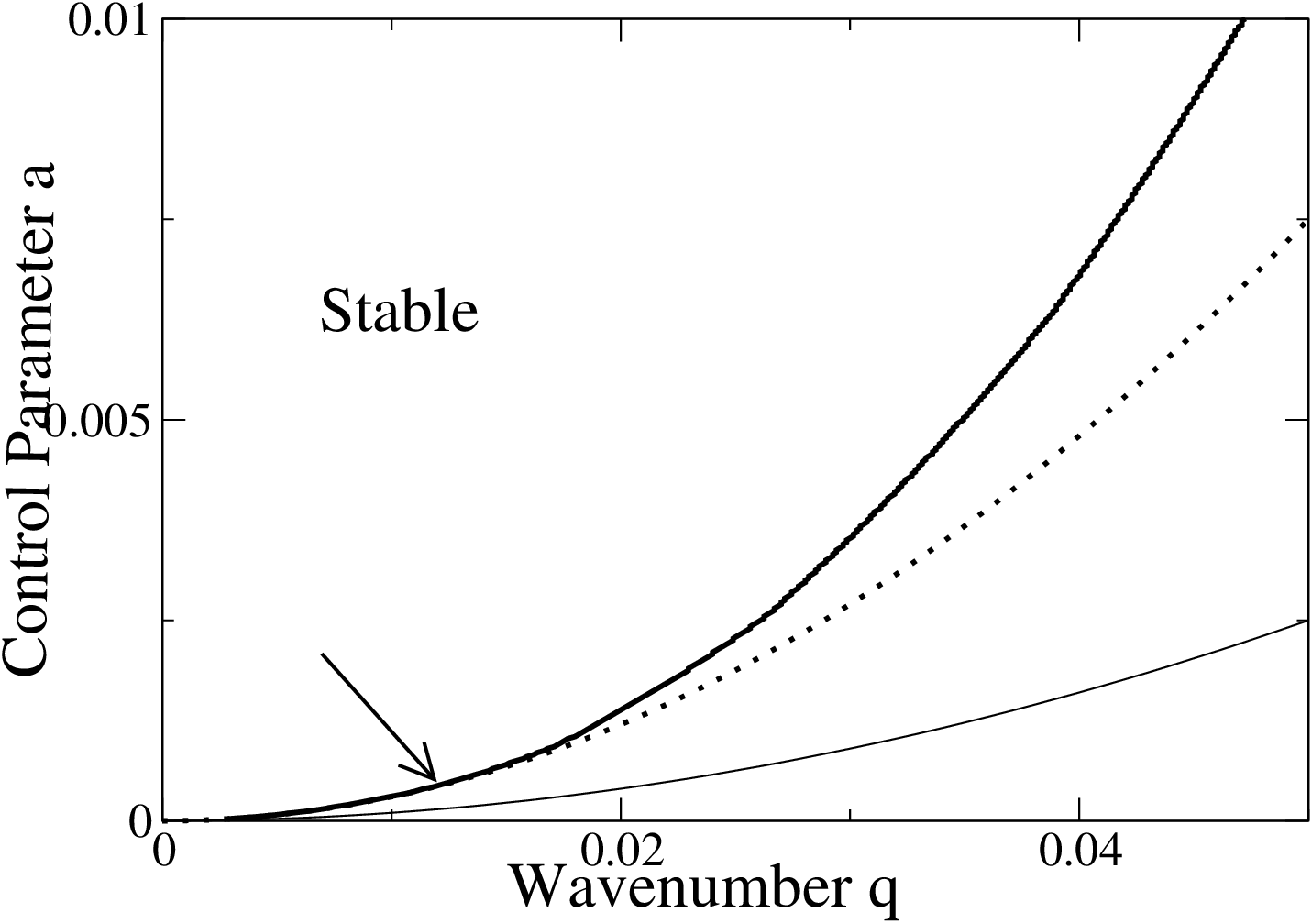}}\hspace{0.2in}\resizebox{3.2in}{!}{\includegraphics{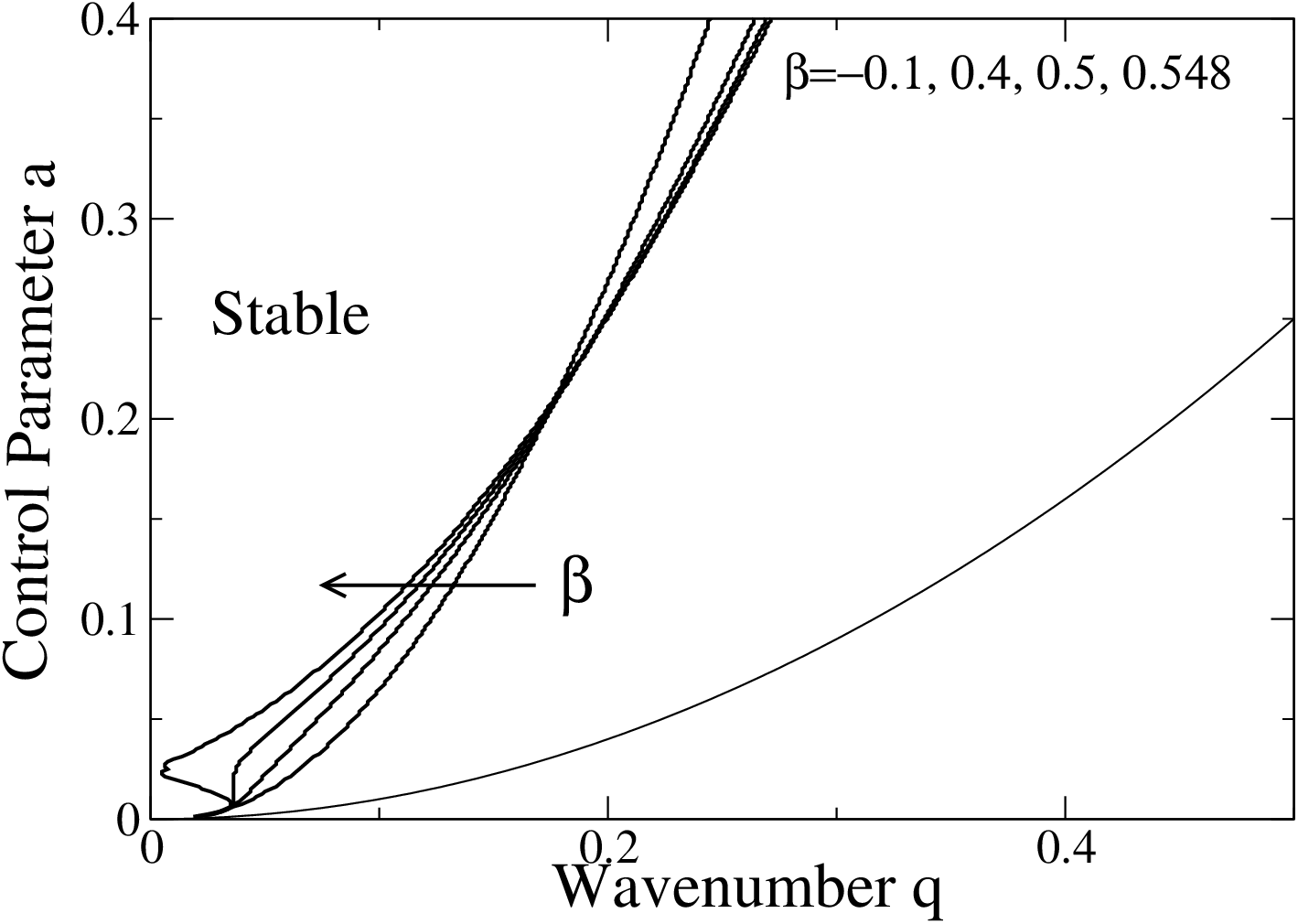}}}
\caption{(a) Linear Stability diagram near threshold of the traveling waves ($a\ll 1$).  The stability boundary approaches the Eckhaus
curve as $a\to 0$ (arrow).
(b) Linear stability diagram with $h$=1.0, $\alpha$=0.02, $\delta$=1.0, $s$=-1.1, -0.6, -0.5, -0.452.  In this intermediate regime both
phase and amplitude instabilities can set in.} \label{Fig:linstabint}
\end{figure}

\section{Phase Instability}

To facilitate the analysis of the phase instability, we make use of the large, negative eigenvalue of the magnitude of \textit{A}
 to derive coupled
phase and field equations in a longwave limit.  If the longwave limit is taken at finite $\alpha$ both the magnitude and the advected field
can be adiabatically eliminated and, to leading order, one obtains the usual phase equation with the usual Eckhaus stability limit.  The
effect of the $C$-field can be captured, however, in a distinguished longwave limit in which $\alpha\to 0$ as well.  Then 
the amplitude can be adiabatically slaved to the dynamics
of the phase and $C$-field, given that we are sufficiently far from the neutral stability boundary ($R^{2}\not\ll 1$).

We consider an ansatz of the form $A=R(T,X)e^{\frac{i}{\epsilon}\phi (T,X)}$, where $X=\epsilon x, T=\epsilon^{2} t$, and additionally
$\alpha = \epsilon^{2}\alpha_{2}$.
 This scaling allows for large variations in the phase, although slow in time and space.  To leading order we obtain an
algebraic relationship relating the slaved amplitude to the C-field and local wavenumber,

\begin{equation}
R^{2}=a+C-(\partial_{X}\phi)^{2}. \label{LW:R}
\end{equation}

Combining equations at orders $\epsilon$ and $\epsilon^{2}$ yields the longwave equations for the phase and the C-field,

\begin{eqnarray}
\partial_{T}\phi &=&D_{E}\partial_{X}^{2}\phi+\frac{\partial_{X}\phi\partial_{X} C}{R^{2}}, \label{LW:phi} \\
\partial_{T}C&=&\delta\partial_{X}^{2}C-\alpha_{2}C+(s+h)\partial_{X}C-h\partial_{X}(\partial_{X}\phi)^{2}. \label{LW:C}
\end{eqnarray} 

The phase-diffusion coefficient $D_{E}$ is defined as before (cf. (\ref{D})) where $R^{2}$ is now given by (\ref{LW:R}) 
and the local wavenumber is now given
by the gradient of the phase, $q = \partial_{X}\phi$.  As before, we linearize the longwave equations about the plane wave state with an
ansatz,

\begin{eqnarray}
\phi &=&qX+\epsilon\tilde{\phi}e^{ipX+\sigma T}, \label{LW:lin:phi}\\
C&=&\epsilon\tilde{C}e^{ipX+\sigma T}. \label{LW:lin:C}
\end{eqnarray}

The linearized longwave equations yield a complex, quadratic dispersion relation.  As for the full EGLE we expand the growth rate $\sigma$ in
powers of the modulation wavenumber $p$,

\begin{eqnarray}
\sigma_{phase}(p)&=&-D_{E}p^{2}+2\frac{ihq^{2}}{R^{2}\alpha_{2}}p^{3}-2\frac{q^{2}\beta}{R^{2}\alpha_{2}^{2}}p^{4}+O(p^{5}), 
\label{LW:eig:phi}\\
\sigma_{field}(p)&=&-\alpha_{2}+i(s+h)p-\delta p^{2}+O(p^{3}). \label{LW:eig:C}
\end{eqnarray}

As expected, phase modes with infinitesimal wavenumber become unstable when $D_{E}$ changes sign.  However, according to (\ref{LW:eig:phi})
 for $\beta < 0$ phase modes with finite wavenumber \textit{p} become unstable already for $D_{E} >0$ if

\begin{equation}
p^{2} > 2D_{E}\frac{R^{2}\alpha_{2}^{2}}{q^{2}\beta}. \label{LW:p}
\end{equation}

Thus, although we have identified the instability in the longwave limit, it is a shortwave instability and to determine
the value of the critical modulation wavenumber p for which $\sigma$ first passes through zero we must solve the dispersion relation exactly.

The action of the linearized operator of the longwave equations on (\ref{LW:lin:phi}, \ref{LW:lin:C})
 yields a 2x2 matrix, the determinant of which must equal zero for a
nontrivial solution to exist.  If we consider the real and imaginary parts of the growth rate $\sigma = \sigma_{r}+i\omega$, this dispersion
relation can be written as,
\begin{eqnarray}
\sigma_{r}^{2}+c_{1}\sigma_{r}+c_{2}&=&0, \label{disp1}\\
c_{3}\sigma_{r}+c_{4}&=&0, \label{disp2}
\end{eqnarray}
where
\begin{eqnarray}
c_{1}&=&(p_{1}^{2}(D_{E}+\delta)+\alpha_{2}), \label{disp:c1}\\
c_{2}&=&-\omega^{2}+\omega p(s+h)+p_{1}^{2}D_{E}(p_{1}^{2}\delta+\alpha_{2}), \label{disp:c2}\\
c_{3}&=&2\omega-p(s+h), \label{disp:c3}\\
 c_{4}&=&\omega (p^{2}(D_{E}+\delta)+\alpha_{2})-\frac{p^{3}}{h}(D_{E}(\beta-h^{2})+h^{2}). \label{disp:c4}
\end{eqnarray}
At criticality ($\sigma_{r}=0$), (\ref{disp2}) and (\ref{disp:c4}) yield the Hopf-frequency,
\begin{equation}
\omega_{Hopf}=\frac{p^{3}}{h}\frac{(D_{E}(\beta -h^{2})+h^{2})}{(p^{2}(D_{E}+\delta)+\alpha_{2})}. \label{freq}
\end{equation}
From equations (\ref{disp1}) and (\ref{disp:c2}) we arrive at a bi-cubic polynomial in p, the roots of which give the 
critical modulation wavenumber for the oscillatory instability,
\begin{eqnarray}
&&-p^{4}(D_{E}(\beta -h^{2})+h^{2})^{2}+p^{2}\beta (D_{E}(\beta -h^{2})+h^{2})(p^{2}(D_{E}+\delta)
+\alpha_{2})+\nonumber\\
&&h^{2}D_{E}(\delta p^{2}+\alpha_{2})(p^{2}(D_{E}+\delta)+\alpha_{2})^{2}=0. \label{poly} 
\end{eqnarray}

 Solving for
p directly in the sixth-order polynomial is not analytically feasible, but we note that the
parameter $\beta$ is merely quadratic in (31) and so seek a solution for $\beta$,
\begin{eqnarray}
\beta_{1,2}&=&-\frac{h^{2}(1-D_{E})(p^{2}(\delta -D_{E})+\alpha_{2})}{2D_{E}(p^{2}\delta +\alpha_{2})}\nonumber\\
&&\pm\frac{p(p^{2}(D_{E}+\delta)+\alpha_{2})}{2D_{E}p^{2}(p^{2}\delta 
+\alpha_{2})}\sqrt{h^{2}p^{2}(1-D_{E})^{2}
-4D_{E}^{2}(p^{2}\delta +\alpha_{2})^{2}}. \label{beta}
\end{eqnarray}
It can be shown that $\sigma_{r} > 0$ only if the real quantity $\beta = h(s+h)$ is in the range $\beta_{1}<\beta <\beta_{2}$.
Thus, the finite-wavenumber instability can only arise if $\beta_{1,2}$ are real, requiring the discriminant to be positive,
$h^{2}p^{2}(1-D_{E})^{2} > 4D_{E}^{2}(p^{2}\delta +\alpha_{2})^{2}$.  This can be interpreted as a constraint
on the modulation wavenumber p,
\begin{equation}
 |p^2-\frac{h^{2}(1-D_{E})^{2}-8D_{E}^{2}\delta\alpha_{2}}{8D_{E}^{2}\delta^{2}}|<\frac{h(1-D_{E})}{8D_{E}^{2}\delta
^{2}}\sqrt{h^{2}(1-D_{E})^{2}-16D_{E}^{2}\delta\alpha_{2}}. \label{discriminant}
\end{equation}
This implies the additional constraint $h^{2}(1-D_{E})^{2}>16D_{E}^{2}\delta\alpha_{2}$ which can be simplified to,
\begin{equation}
q^{2} > \frac{a}{3+\frac{|h|}{2\sqrt{\alpha_{2}\delta}}}. \label{lbound}
\end{equation}
This somewhat intricate analysis of (\ref{poly}) can now be summarized concisely.  A \emph{sufficient} condition for stability
of the waves with respect to shortwave instabilities is that the underlying wavenumber of the waves be in the 
band given by $q^{2} < \frac{a}{3+\frac{|h|}{2\sqrt{\alpha_{2}\delta}}}$, which is narrowed with respect to the 
traditional Eckhaus band.  Thus, with respect to phase instabilities,
 the band of stable wavenumbers is bounded below by (\ref{lbound}) and above by the Eckhaus curve,
\begin{equation}
\frac{a}{3+\frac{|h|}{2\sqrt{\alpha_{2}\delta}}} < q_{phase}^{2} < \frac{a}{3}. \label{qbounds}
\end{equation}
If (\ref{qbounds}) is satisfied there exists a range in $\beta$ given by (\ref{beta}) for which the plane wave is unstable.  
For a given $\beta$ from this range the destabilizing modulation wavenumbers are given by the condition that the associated interval 
$\lbrack\beta_{1},\beta_{2}\rbrack$ include that value of $\beta$.

\begin{figure}[t]
\centerline{\resizebox{3.5in}{!}{\includegraphics{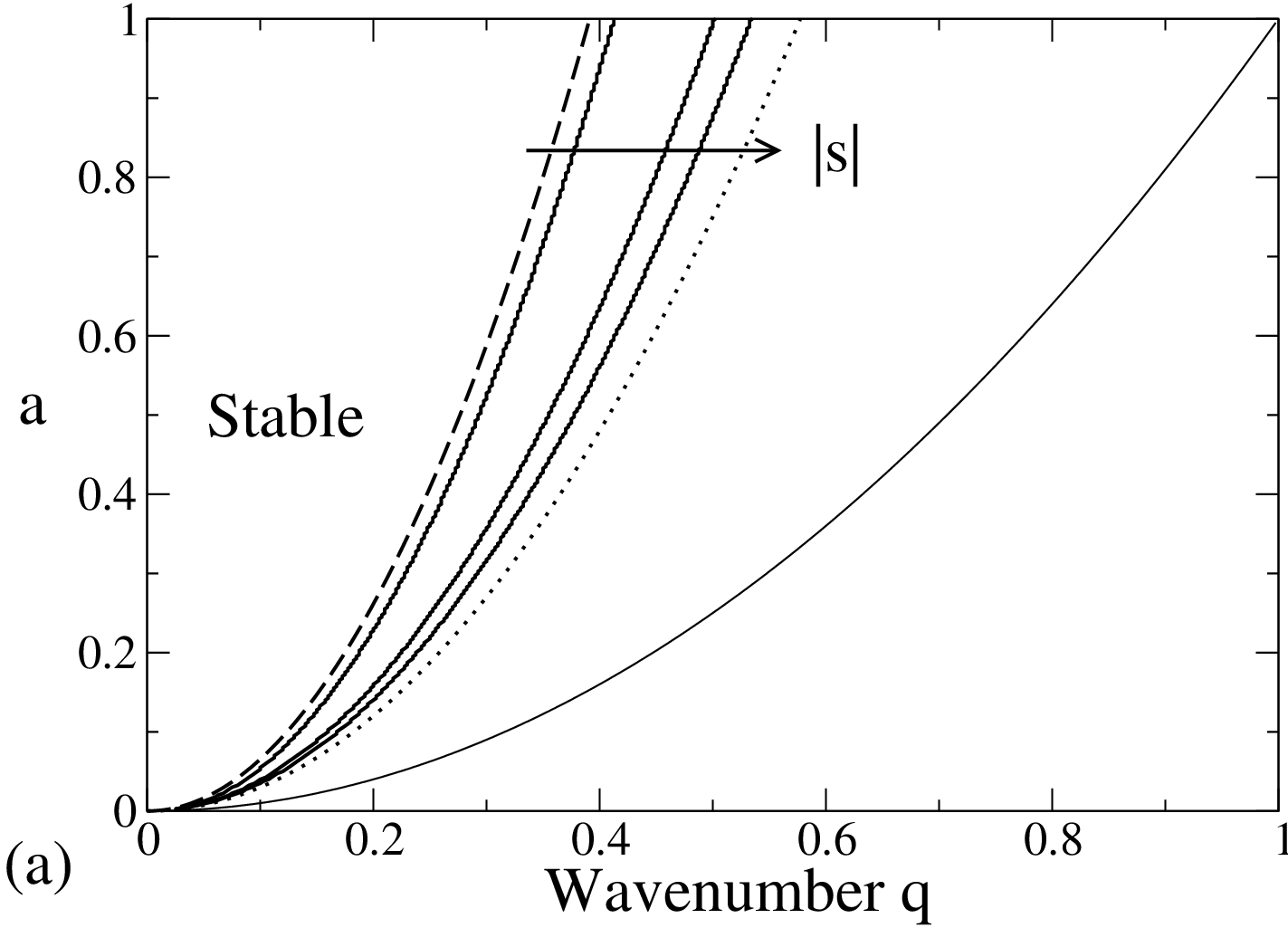}}\hspace{0.2in}\resizebox{3.5in}{!}{\includegraphics{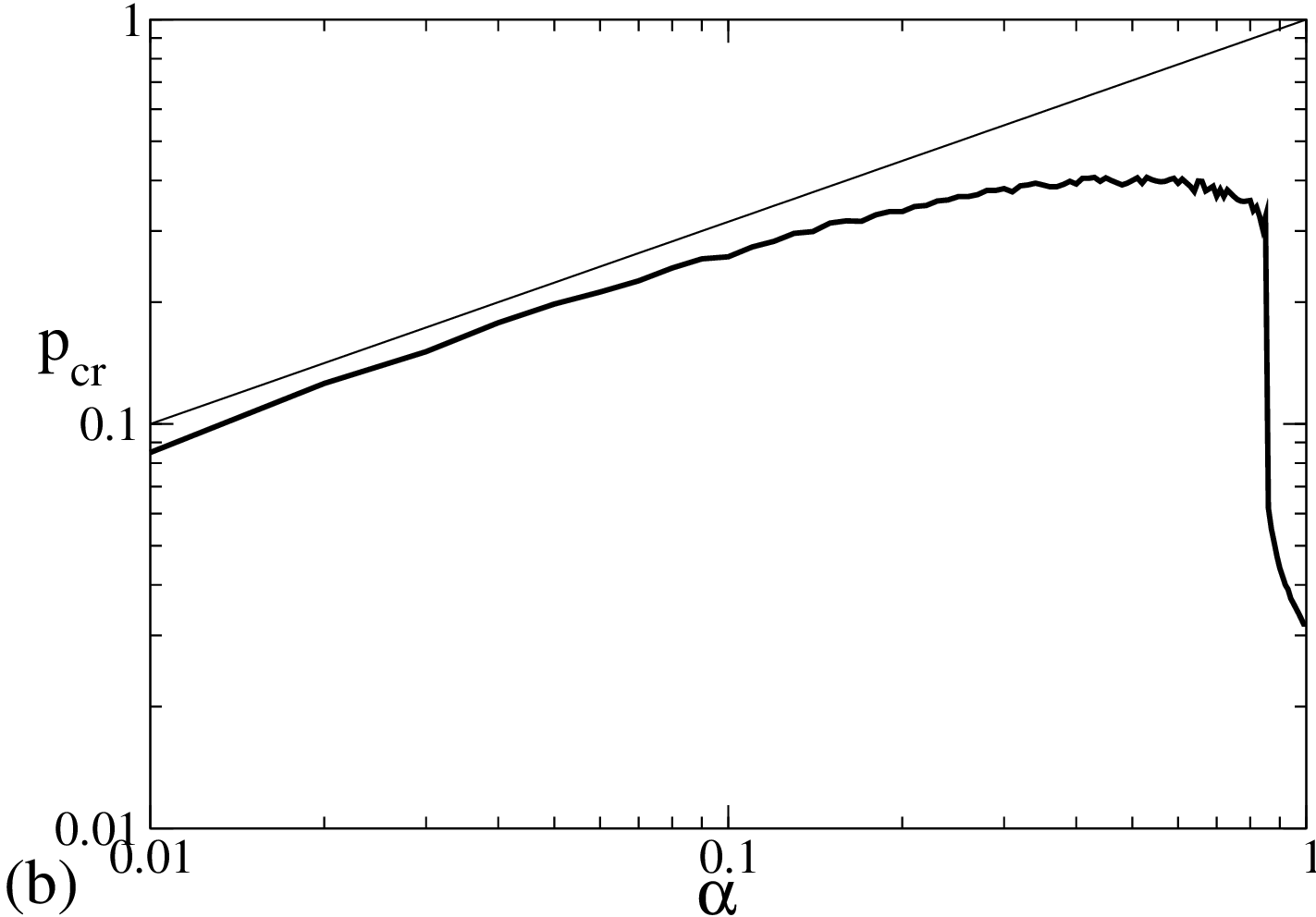}}}
\caption{(a) Stability boundary as a function of the group velocity s.  The boundary is bounded to the left by the stability condition
derived from the dispersion relation of the longwave equations and to the right by the Eckhaus curve. Here h = 1, $\alpha$ = 0.02
$\delta$ = 1.0 and s = -1.5, -3.0, -5.0. (b) Dependence of the critical modulation wavenumber on the decay rate of C, $\alpha$. 
 The thick line is numerically calculated from the full dispersion relation while the thinner line is p=$\sqrt{\frac{\alpha}{\delta}}$.  
(\ref{pcr}) is thus valid only for small enough $\alpha$. ($a$=1.0, $h$=1.0, $s$=-1.5, $\delta$=1.0)} \label{Fig:linstab:pdepen}
\end{figure}

In the limit of zero coupling ($h\to 0$) or large decay rate of the concentration field
($\alpha_{2}\to\infty$) the Eckhaus band is recovered in a consistent manner.  
A linear stability diagram of waves as obtained from the full EGLE (\ref{A}, \ref{C}) is shown in Figure \ref{Fig:linstab:pdepen}(a) 
for various values of the group velocity s.  It can be seen that inequality (\ref{qbounds}) is satisfied.

\subsection{Weakly Nonlinear Analysis of Phase Instability}

To examine the weakly-nonlinear behavior of the phase-instability one must solve (\ref{poly}) for given values of the system
parameters.  In the general case this yields a numerical value for the critical modulation wavenumber $p$ which can then be 
used in a weakly-nonlinear analysis.  The coefficients of the resulting amplitude equation must then be evaluated  numerically.
 However, if we restrict the wavenumber of the waves to lie on the curve 
representing the lower bound of existence of the shortwave instability, we
can obtain analytical values for the critical modulation wavenumber and Hopf-frequency and consequently also for the coefficients of the 
amplitude equation.  Along that curve,

\begin{eqnarray}
q_{cr}^{2}&=&\frac{a_{0}}{3+\frac{|h|}{2\sqrt{\alpha_{2}\delta}}}, \label{qcr}\\
p_{cr}^{2}&=&\frac{\alpha_{2}}{\delta}, \label{pcr}\\
\omega_{cr}&=&-\textrm{sgn}(p_{cr}h)\frac{p_{cr}^{2}}{1+\frac{4\sqrt{\alpha_{2}\delta}}{|h|}}, \label{freqcr}\\
\beta_{cr}&=&-4p_{cr}\delta\left( 1-\frac{1}{2\delta (1+\frac{4\sqrt{\alpha\delta}}{|h|})}\right). \label{betacr}
\end{eqnarray}

It is good to keep in mind that these values fix a unique value of the parameter $\beta = h(s+h)$ given by (\ref{betacr}) 
and are in this sense restrictive.
  In addition, (\ref{qcr}-\ref{freqcr}) are obtained in a longwave analysis, which requires that $p, \alpha\ll 1$. Figure \ref{Fig:linstab:pdepen}(b)
indicates the values of $\alpha$ for which (\ref{qcr}-\ref{freqcr}) are valid.

  We expand the phase and $C$-field in small-amplitude, normal modes, where $\epsilon = \sqrt{\frac{(a-a_{0})}{a_{2}}}$
 is a measure of the distance from the bifurcation point ($a_{0}, q_{cr}$),

\begin{figure}[t]
\centerline{\resizebox{5.0in}{!}{\includegraphics{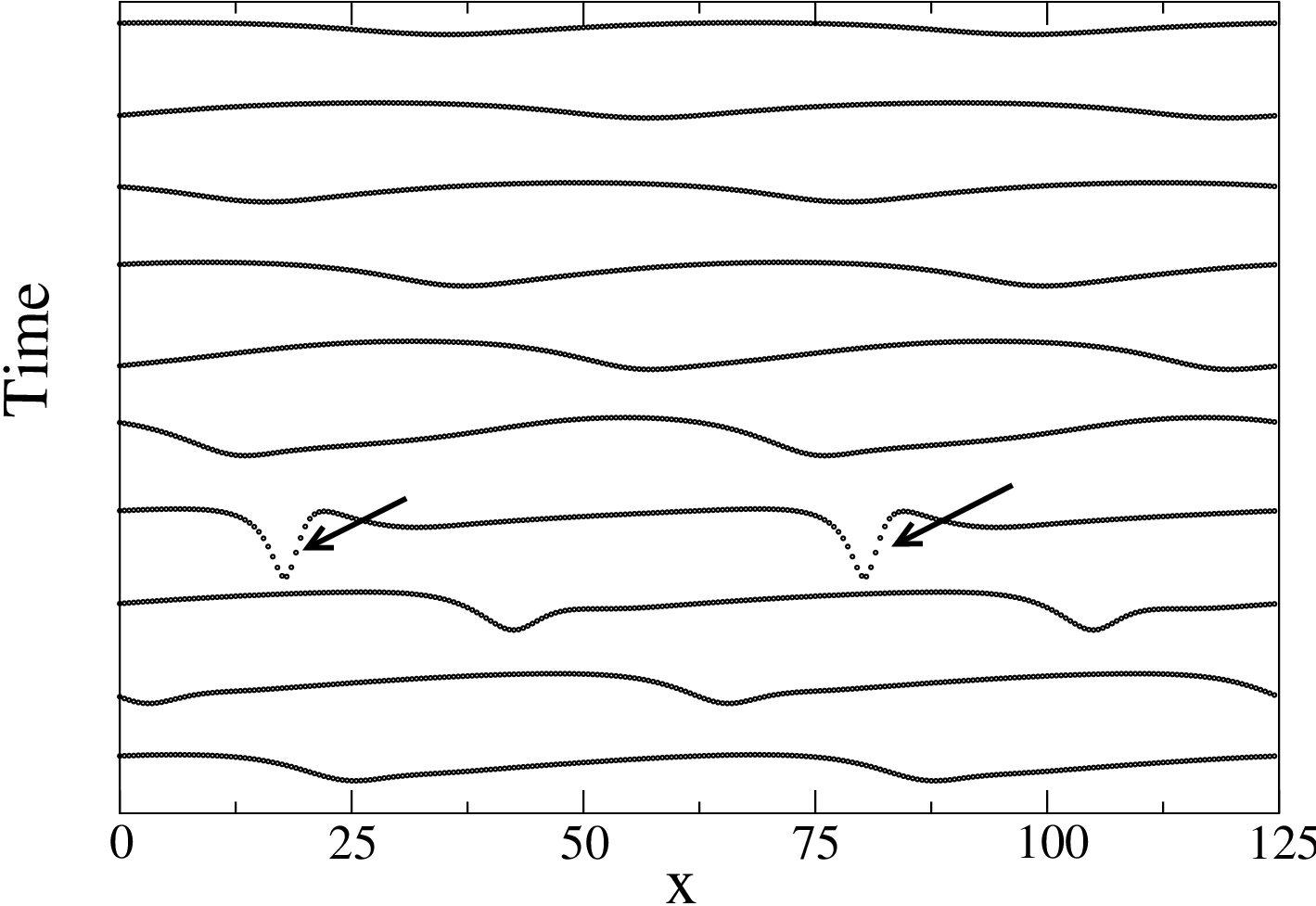}}}
\caption{Space-time diagram of the evolution of the magnitude $|A|$ leading  to a phase slip (arrows), here indicated by the dip in amplitude. 
s=-1.5, h=1
, $\alpha$ =0.02, $\delta$ =1.0} \label{Fig:phaseslip}
\end{figure}

\begin{equation}
\left( \begin{array}{cc}
\phi \\  C \end{array} \right)=
\left( \begin{array}{cc}
q_{cr}X \\ 0 \end{array} \right)+\epsilon\Bigg[
\left( \begin{array}{cc}
\Phi_{0} \\ C_{0} \end{array} \right) A_{0}(\tau )e^{i(p_{cr}X+\omega_{cr}T)}+c.c+h.o.t.\Bigg]. \label{LW:WNL:ansatz}
\end{equation}

The complex amplitude $A_{0}$ of the unstable mode evolves on the superslow time scale $\tau =\epsilon^{2}T$.  
Inserting this ansatz into (\ref{LW:phi}, \ref{LW:C})
 and solving order by order leads to a solvability condition at O($\epsilon^{3}$) which yields a differential equation for $A_{0}$, 

\begin{equation}
\partial_{\tau}A_{0}=-\Sigma (a_{0},\alpha_{2},\delta,h)a_{2}A_{0}+\Gamma (a_{0},\alpha_{2},\delta,h)|A_{0}|^{2}A_{0}. \label{LW:ampeq}
\end{equation}

We are interested in the real part of the cubic coefficient, the sign of which gives the nature of the Hopf-bifurcation 
(forward/backward).
  The coefficient $\Gamma$ is a complicated expression of $\delta$, $\alpha_{2}$, \textit{a}, and \textit{h}, but we 
can expand it for small $\alpha_{2}$ to obtain an
asymptotic approximation ($\alpha$ must be small from Figure \ref{Fig:linstab:pdepen}(b)),  
\begin{eqnarray}
\Sigma &=& \frac{1}{(2\delta +1)}\Bigg(\frac{2|h|}{a_{0}^{3}\delta^{1/2}}\Bigg)^{1/2}(1+i\textrm{sgn}(h))\alpha_{2}^{3/4}+O(\alpha^{5/4}),
\label{Sigma}\\
\Gamma &=& \frac{1}{a^{2}\delta^{3/2}(2\delta +1)}\Bigg(\frac{404}{5|h|}\alpha_{2}^{1/2}+i\textrm{sgn}(h)\Bigg[\frac{12}{\delta^{1/2}}+
\frac{4}{5}\frac{
(494\delta +451)}{(2\delta +1)}\alpha_{2}^{1/2}\Bigg]\Bigg)\alpha_{2}^{3}+O(\alpha_{2}^{9/4}). \label{Gamma}
\end{eqnarray}
Thus in the limit $\alpha_{2}\to 0$, $\Gamma > 0$ for all values of the coefficients, and the phase mode undergoes a backward Hopf-bifurcation.   
Solving the coefficient $\Gamma$ numerically over a wide range of parameter values always yielded a positive real part. 
This analysis does not indicate whether or not the instability saturates at higher order or if the phase becomes undefined, signaling a phase
slip.  To investigate the nonlinear behavior of the phase, (\ref{A}, \ref{C}) were integrated numerically using a linearized Crank-Nicholson scheme.  It was
found for the parameter values tested, that the waves outside the stability band underwent a phase slip when perturbed, relaxing to the plane-wave-
stable band as seen in Figure \ref{Fig:phaseslip}, which shows a space-time diagram of the magnitude $|A|$ of the waves.

\section{Amplitude Instability: Modulated Waves}

\begin{figure}[t]
\centerline{\resizebox{3.5in}{!}{\includegraphics{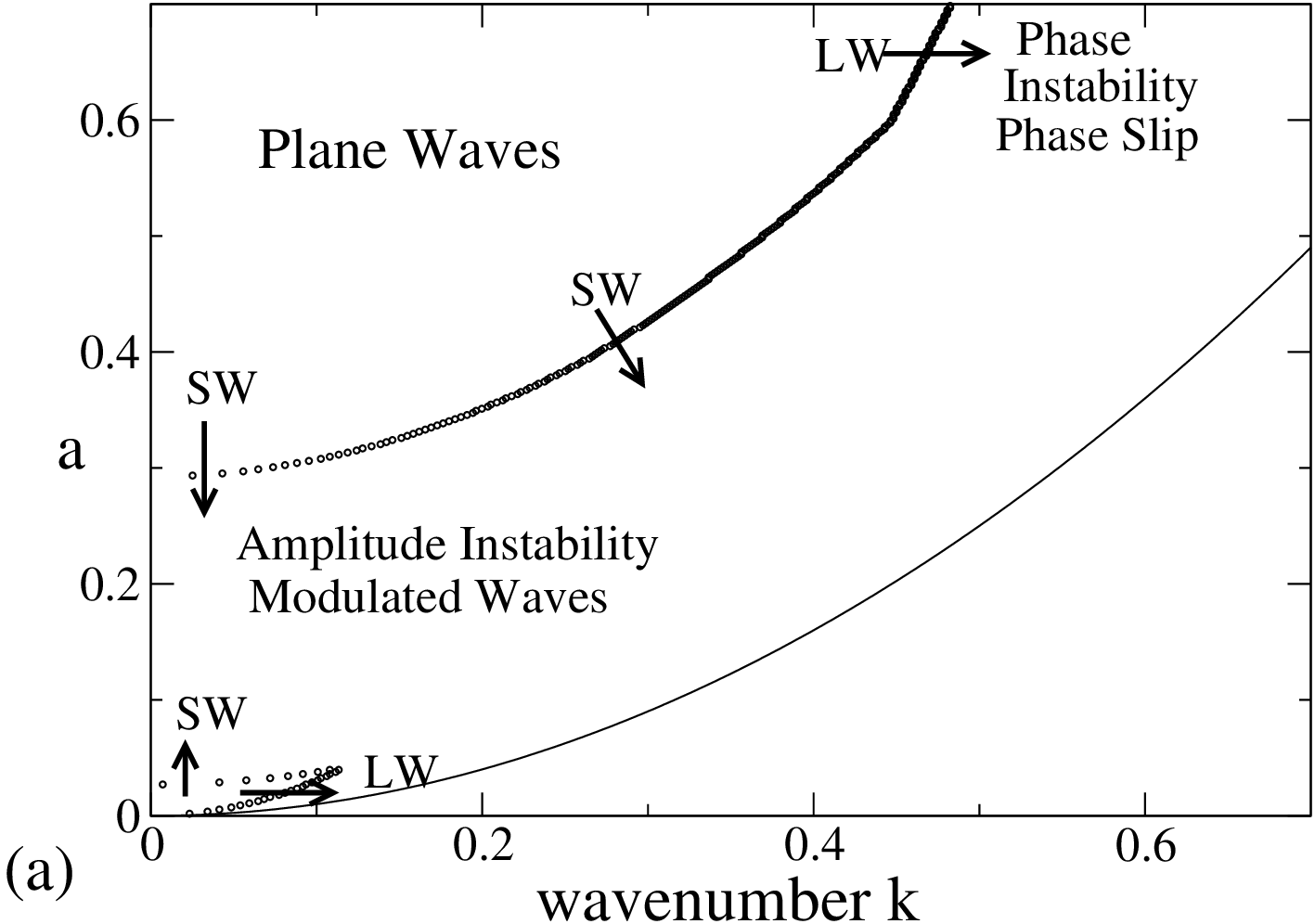}}\hspace{0.2in}\resizebox{3.5in}{!}{\includegraphics{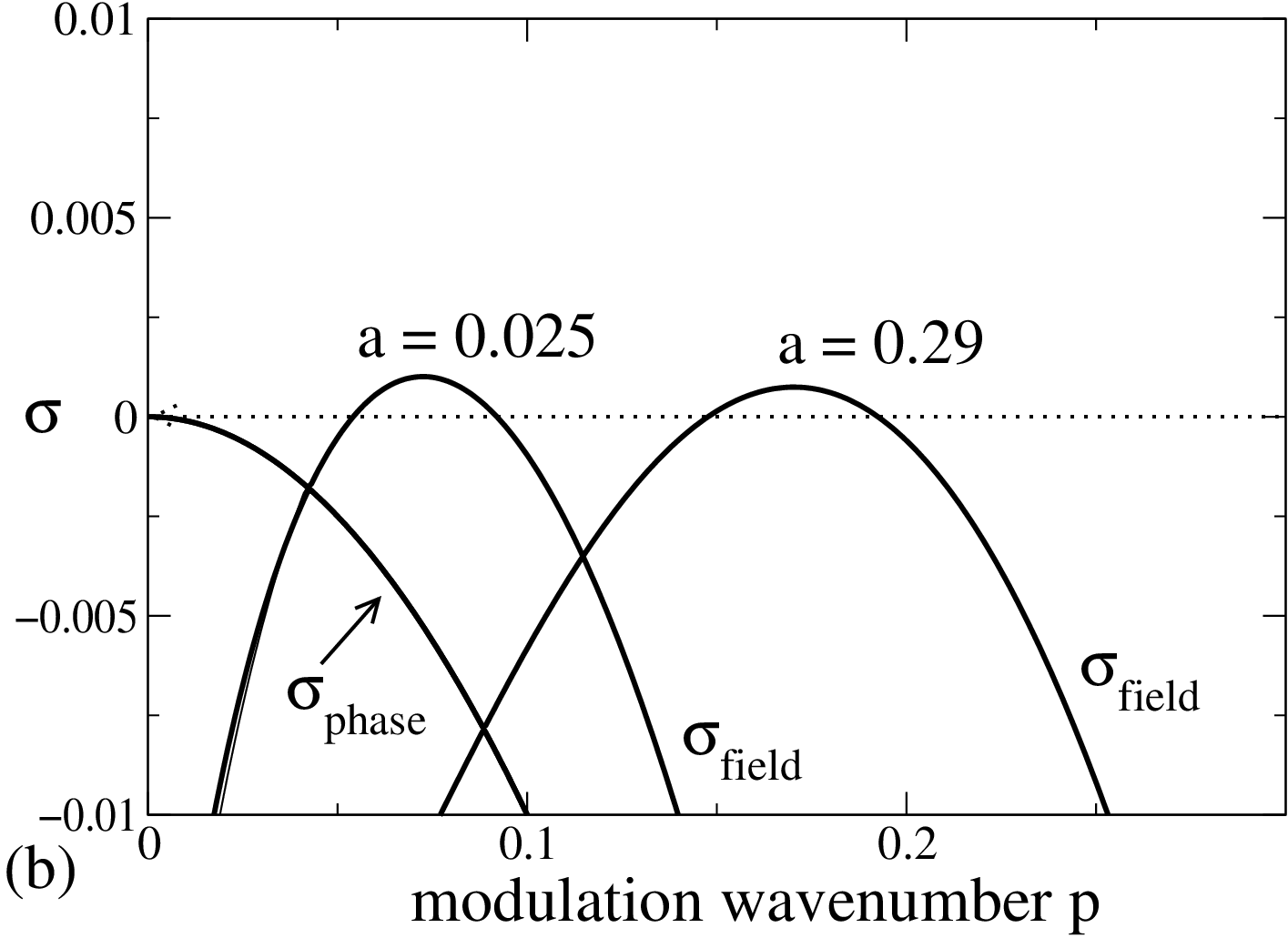}}}
\caption{A linear stability diagram is given in (a) for h=1, s=1, $\alpha$=0.02, $\delta$=1.5. LW and SW denote longwave and shortwave
instabilities respectively.  In (b) the largest positive eigenvalue of the system is plotted versus the modulation wavenumber at bandcenter.
 Shortwave instabilities occur as the control parameter a is increased from the stable region near onset of the plane waves
and as a is decreased from the stable region far beyond threshold.} \label{Fig:linstabamp}
\end{figure}

We now investigate the instability corresponding to the eigenvalue of the $C$-field passing through zero in (\ref{eig:Cbc}) which occurs only for $\beta >0$.  
Figure \ref{Fig:linstabamp}(a) indicates that plane waves are stable at onset, becoming linearly unstable as the control parameter is increased until
, for large enough values of the control parameter, they once again regain stability.  In addition, the band of stable wavenumbers is bounded 
by the Eckhaus curve outside the range of the amplitude instability.  Thus for $\beta >0$ two distinct instabilities are possible depending on the 
wavenumber of the waves and the distance from threshold.
  Figure \ref{Fig:linstabamp}(b) shows the relevant eigenvalue as a function of 
the modulation wavenumber for increasing values of the control parameter.  The wavenumber of the fastest growing mode increases continuously over the
range of values of the control parameter for which the waves are linearly unstable.

At bandcenter, the system (\ref{EGLE:R}, \ref{EGLE:theta}, \ref{EGLE:C2}) decouples in the case of unperturbed plane waves. 
 Perturbing the plane-wave solution with the ansatz (\ref{lin:R}, \ref{lin:theta}, \ref{lin:C}) where q=0, 
introduces an order $\epsilon^{2}$ coupling of the phase to the real amplitude through $(\partial_{x}\theta)^{2}$,
 which does not appear in the linearized system.  Thus it is sufficient to consider the following equations for the linear stability of
the waves at bandcenter,
 
\begin{eqnarray}
\partial_{t}R&=&\partial_{x}^{2}R+R[(a+C)-R^{2}], \label{BC:R}\\
\partial_{t}C&=&\delta\partial_{x}^{2}C-\alpha C+s\partial_{x}C+2hR\partial_{x}R. \label{BC:C}
\end{eqnarray}

Linearizing about the plane-wave solution in the above equations yields, for spatially periodic modulations, a complex, quadratic polynomial in the 
growth rate $\sigma$.  As in section 3, setting $\sigma_{r}=0$ gives a value for the Hopf-frequency,

\begin{equation}
\omega_{Hopf} = p\frac{(2a_{0}(s+h)+sp^{2})}{((2a_{0}+\alpha )+(1+\delta)p^{2})}, \label{BC:freq}
\end{equation}
and a polynomial in p, the roots of which give the values of the critical modulation wavenumber,

\begin{eqnarray}
-p^{2}(2a_{0}(s+h)+sp^{2})^{2}+sp^{2}(2a_{0}(s+h)+sp^{2})((2a_{0}+\alpha )+(1+\delta )p^{2})\nonumber\\
+(p^{2}+2a_{0})(\delta p^{2}+\alpha)((2a_{0}+\alpha )+(1+\delta )p^{2})^{2}=0. \label{BC:poly}
\end{eqnarray}

If we are to solve (\ref{EGLE:R}),(\ref{EGLE:theta}),(\ref{EGLE:C2}) to higher order, we must consider the coupling of the 
phase to the amplitude.  We take an expansion of the form,

\begin{equation}
\left( \begin{array}{ccc}
R \\ \theta \\ C \end{array}\right)=
\left( \begin{array}{ccc}
\sqrt{a_{0}} \\ 0 \\ 0 \end{array}\right)+\epsilon
\left( \begin{array}{ccc}
R_{0} \\ \theta_{0} \\ C_{0} \end{array}\right)+\epsilon^{2}
\left( \begin{array}{ccc}
R_{1} \\ \theta_{1} \\ C_{1} \end{array}\right)+\dots \label{BC:expand}
\end{equation}
Plugging this into (\ref{EGLE:theta}) yields the following series of equations for the phase,

\begin{eqnarray}
\partial_{t}\theta_{0}&=&\partial_{x}^{2}\theta_{0}, \label{BC:theta0}\\
\partial_{t}\theta_{k}&=&\partial_{x}^{2}\theta_{k}+\partial_{x}\Bigg(\sum_{n=1}^{k-1}\frac{1}{n!}\frac{\partial^{n}(R^{2})}
{\partial\epsilon^{n}}(\epsilon =0)\epsilon^{n}\Bigg)\partial_{x}\theta_{j}. \label{BC:thetak}
\end{eqnarray}

Where n+j=k and k$>$0.  Thus to leading order the phase satisfies the constant coefficient diffusion equation.  Higher orders include
a forcing term proportional to the gradient of all lower order terms which must all decay to zero for long times due to (\ref{BC:theta0}).  We conclude
that the long-term behavior of small-amplitude instabilities at bandcenter will not be affected by the phase and consider (\ref{BC:R}, \ref{BC:C})
the relevant dynamics.
\subsection{Weakly Nonlinear Analysis}

In order to capture the dynamics of the instability near onset at both extremes of the unstable band in the control parameter, we 
carry out a weakly nonlinear analysis.  Since solving the polynomial (\ref{BC:poly}) for the critical modulation wavenumber is intractable
we leave \textit{p} as an implicit variable.

We again introduce a slow time scale $\tau =\epsilon^{2}t$ and define the distance from onset $\epsilon =\sqrt{\frac{(a-a_{0})}{a_{2}}}$.  
We consider a small-amplitude expansion,
\begin{displaymath}
\left( \begin{array}{cc}
R \\ C \end{array} \right) =
\left( \begin{array}{cc}
\sqrt{a_{0}} \\ 0 \end{array} \right)+\epsilon\Bigg[
\left( \begin{array}{cc}
R_{0} \\ C_{0} \end{array} \right) A_{0}(\tau )e^{i(p_{cr}x+\omega_{cr}t)}+c.c+h.o.t.\Bigg],
\end{displaymath}  \label{BC:ansatz}
\begin{figure}[t!]
\centerline{\resizebox{3.2in}{!}{\includegraphics{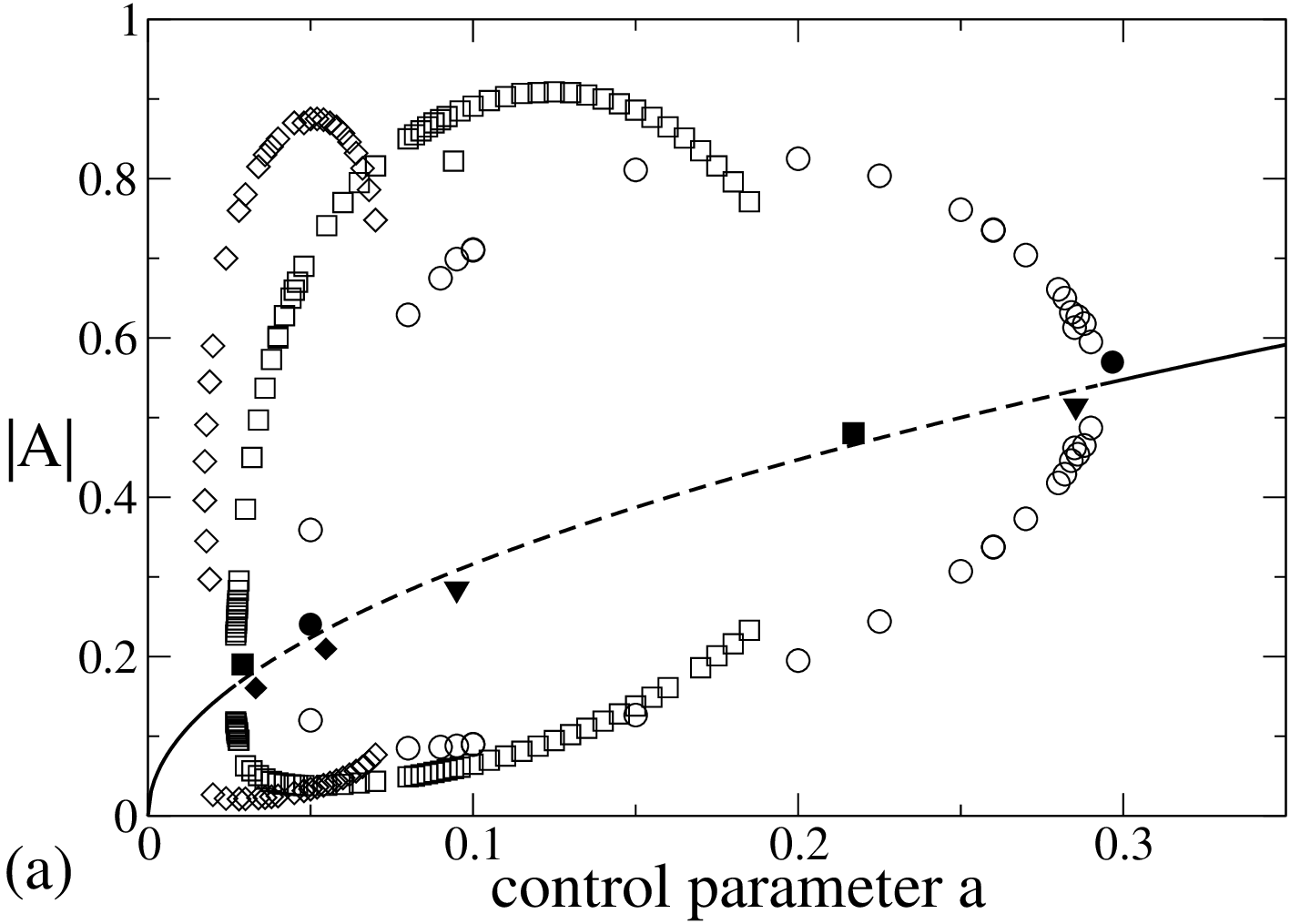}}\hspace{0.2in}\resizebox{3.2in}{!}{\includegraphics{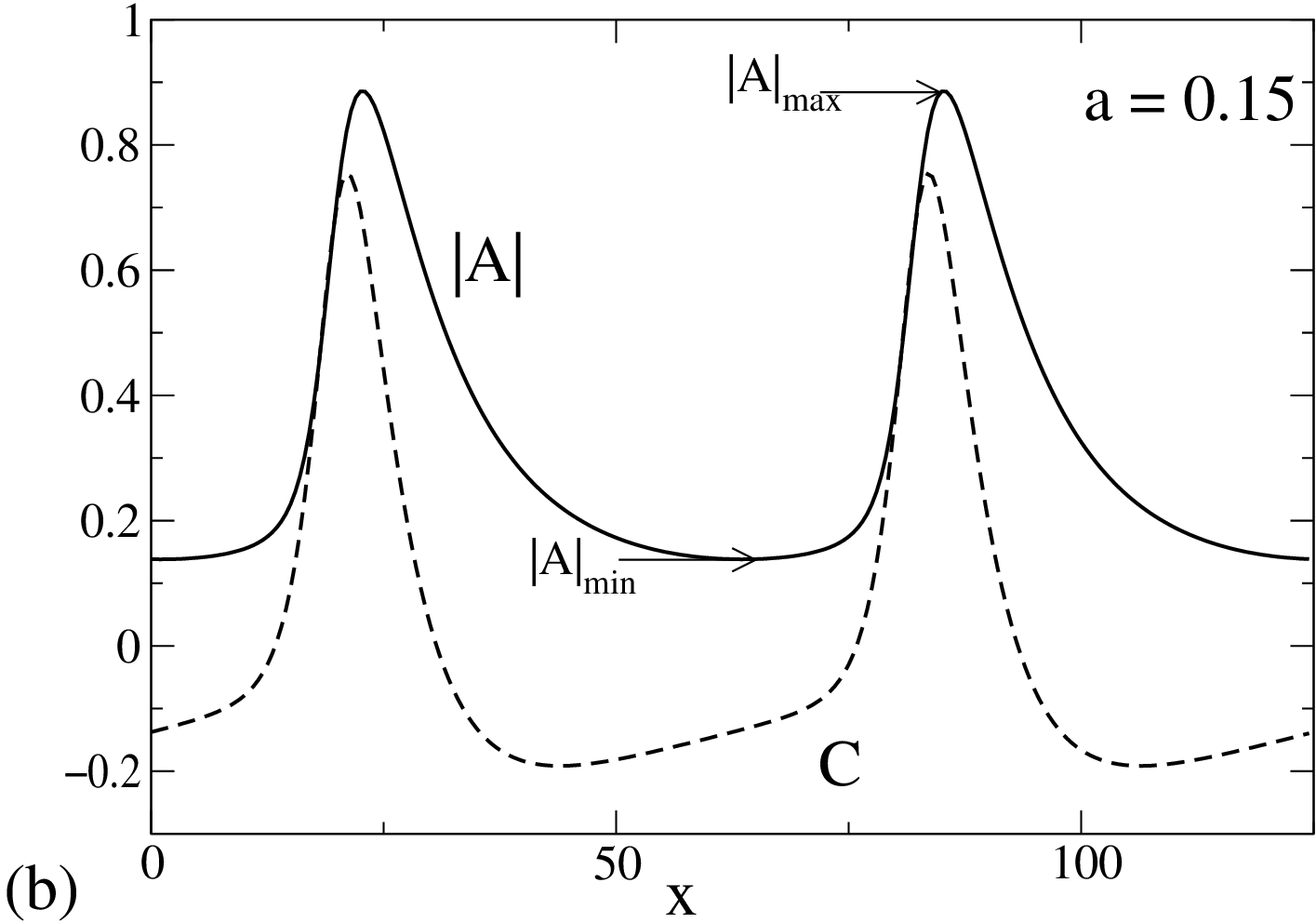}}}
\vspace{0.4in}\resizebox{3.2in}{!}{\includegraphics{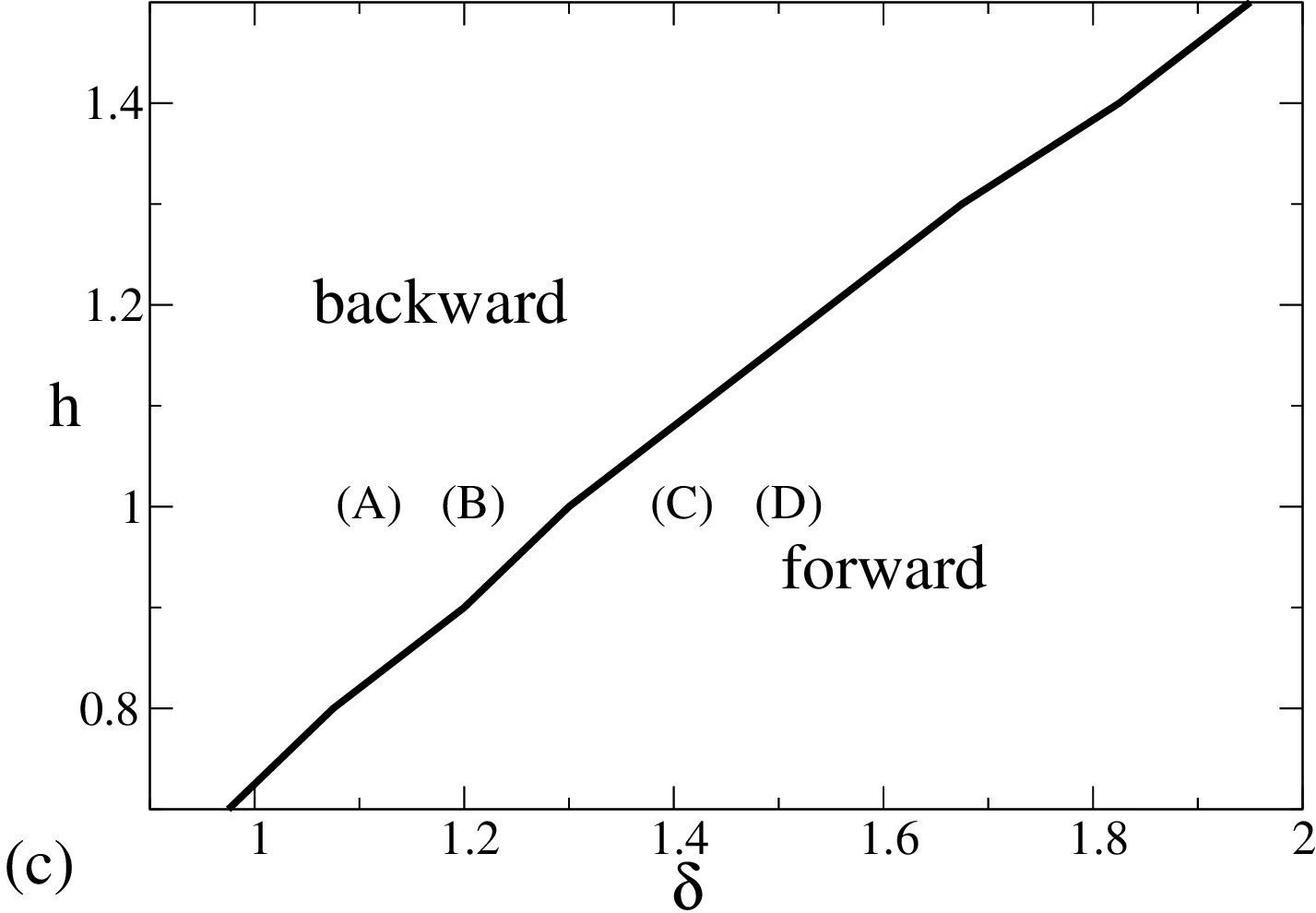}}\hspace{0.2in}\resizebox{3.2in}{!}{\includegraphics{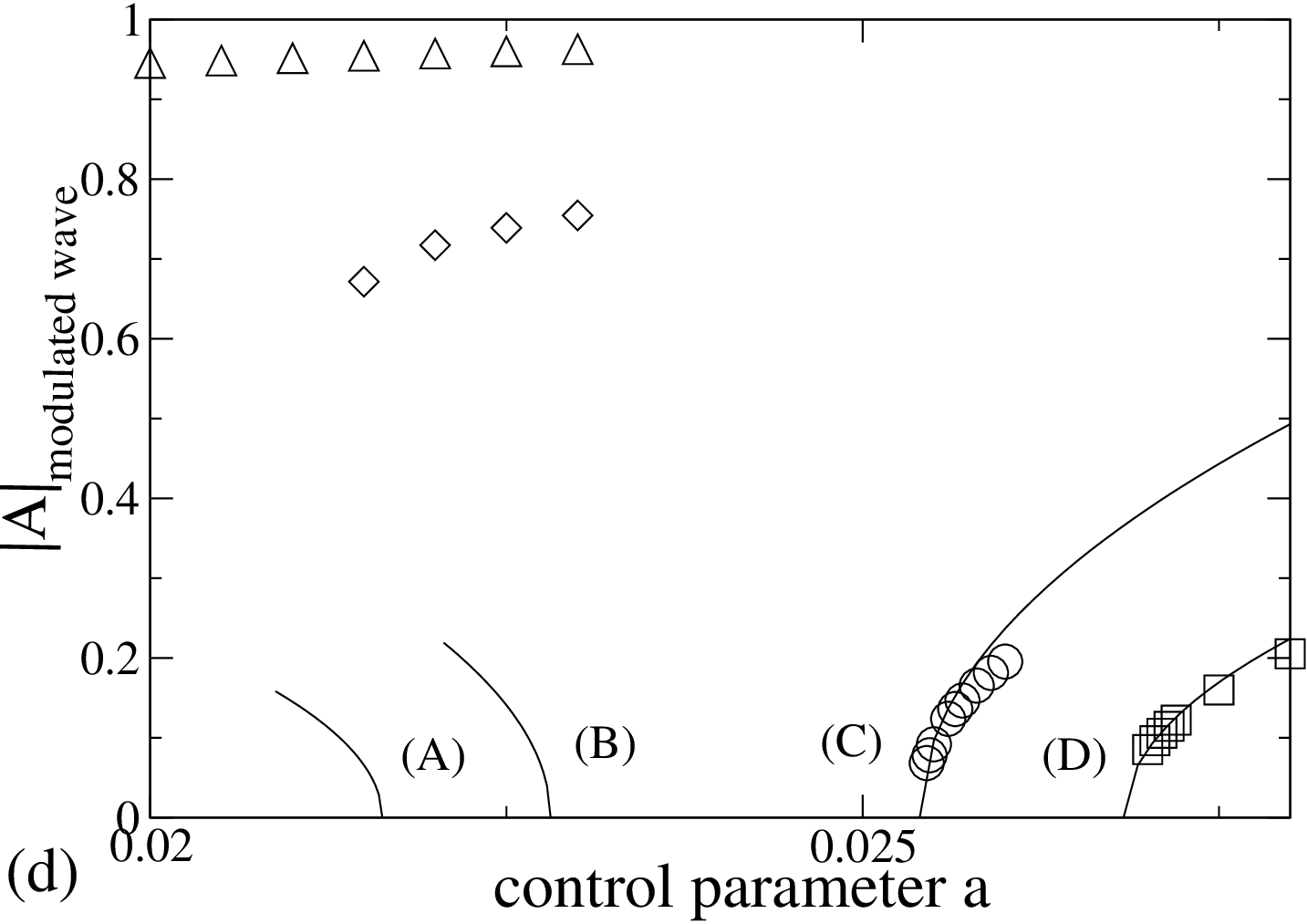}}
\caption{(a) Bifurcation diagram of modulated waves.  Diamonds indicate minima and maxima of a single-wavelength modulation, squares and circles two
and three wavelengths respectively.  Solid symbols indicate the point at which the corresponding mode linearly 
destabilizes the traveling wave (triangles indicate
four-wavelength instability although solution branches are not shown).  Parameter values are s=1, h=1, $\delta$=1.5, $\alpha$=0.02 with a 
system size L=125. (b) Solution with 2-wavelength modulation.
(c) Switch of the bifurcation from backward to forward according to (\ref{BC:ampeq}). (d) Comparison of
the results from the amplitude equation (solid lines) to numerical simulations confirms this transition.  Enough data points
where taken for (A),(B) to indicate the presence of bistability and thus confirm the subcritical nature of the bifurcations.
s=1, h=1, $\alpha$=0.02, and $\delta$=1.1 (triangles),1.2 (diamonds),1.4 (circles),1.5 (squares) with a system size L=62.5.} \label{Fig:bifmodwave}
\end{figure}
and systematically arrive at a solvability condition at order O($\epsilon^{3}$) which yields a differential equation for $A_{0}$,  
\begin{equation}
\partial_{\tau}A_{0}=\Lambda (a_{0},\alpha,\delta,h,s;p(a_{0},..))a_{2}A_{0}+\Pi (a_{0},\alpha,\delta,h,s;p(a_{0},..))|A_{0}|^{2}A_{0}.
\label{BC:ampeq}
\end{equation}

\begin{figure}[t!]
\centerline{\resizebox{3.2in}{!}{\includegraphics{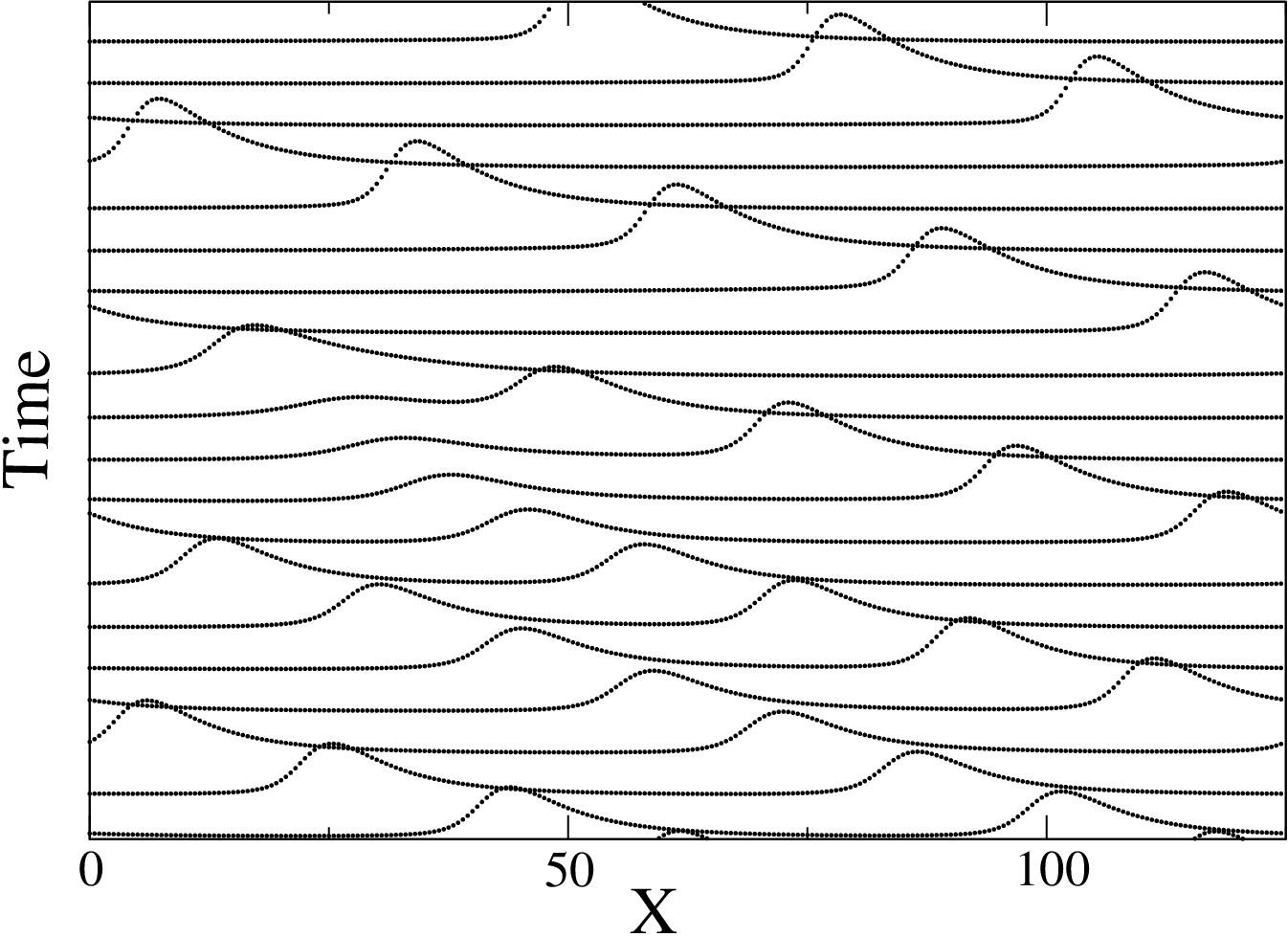}}}
\caption{Space-time diagram of $|A|$ showing the annihilation of one 'hump' by another leading to the formation of a single-wavelength 
modulated pattern from a two-wavelength
pattern.  s=1, h=1, $\alpha$=0.02, $\delta$=1.5, a=0.04} \label{Fig:2to1}
\end{figure}

The coefficients $\Lambda$ and $\Pi$ in (\ref{BC:ampeq}) 
must be solved numerically by determining the value of p from (\ref{BC:poly}) for given values of the parameters.  For the values of the parameters
tested, it was found that the secondary bifurcation encountered first as the control parameter $a$ is increased from 0  can be either
forward or backward (cf. stability island including \textit{a}=0 in Fig. \ref{Fig:linstabamp}(a)).
  The bifurcation which occurs as $a$ is decreased from the region of stability of the plane waves 
was, in all cases, supercritical.  A bifurcation diagram for the modulated waves is given in Figure \ref{Fig:bifmodwave}(a) for a system size of L=125.  
The particular structure of the cascade of bifurcations is a consequence of the finite system size, where each branch corresponds to 
a discrete number of wavelengths
of the modulation.  The bifurcation to modulated waves near onset of the traveling waves (for small \textit{a}) is examined in detail in Figures
\ref{Fig:bifmodwave}(c),(d) where the amplitude equation is compared to numerical simulation.  As $\delta$ is decreased for fixed values of the other
parameters, the bifurcation switches from forward to backward.  Figure \ref{Fig:bifmodwave}(d) confirms the validity of the amplitude equation.

The dynamics of the modulated wave pattern is not trivial.  The destruction or creation of a modulation wavelength during the transient
dynamics before the stable pattern is achieved occurs due to particle-like interactions of the `humps' of the modulations as seen in
Figure \ref{Fig:2to1}.  The bifurcation diagram in Figure \ref{Fig:bifmodwave}(a) indicates the existence of multiple branches for a single value of the control
parameter.  If the initial bifurcation is to a lower branch (say with 3 wavelengths), two of the three 'humps' will collide after 
an initial periodically modulated state.  The remaining two may again collide to form a single-wavelength modulation which is then
stable.

\begin{figure}[t!]
\centerline{\resizebox{3.2in}{!}{\includegraphics{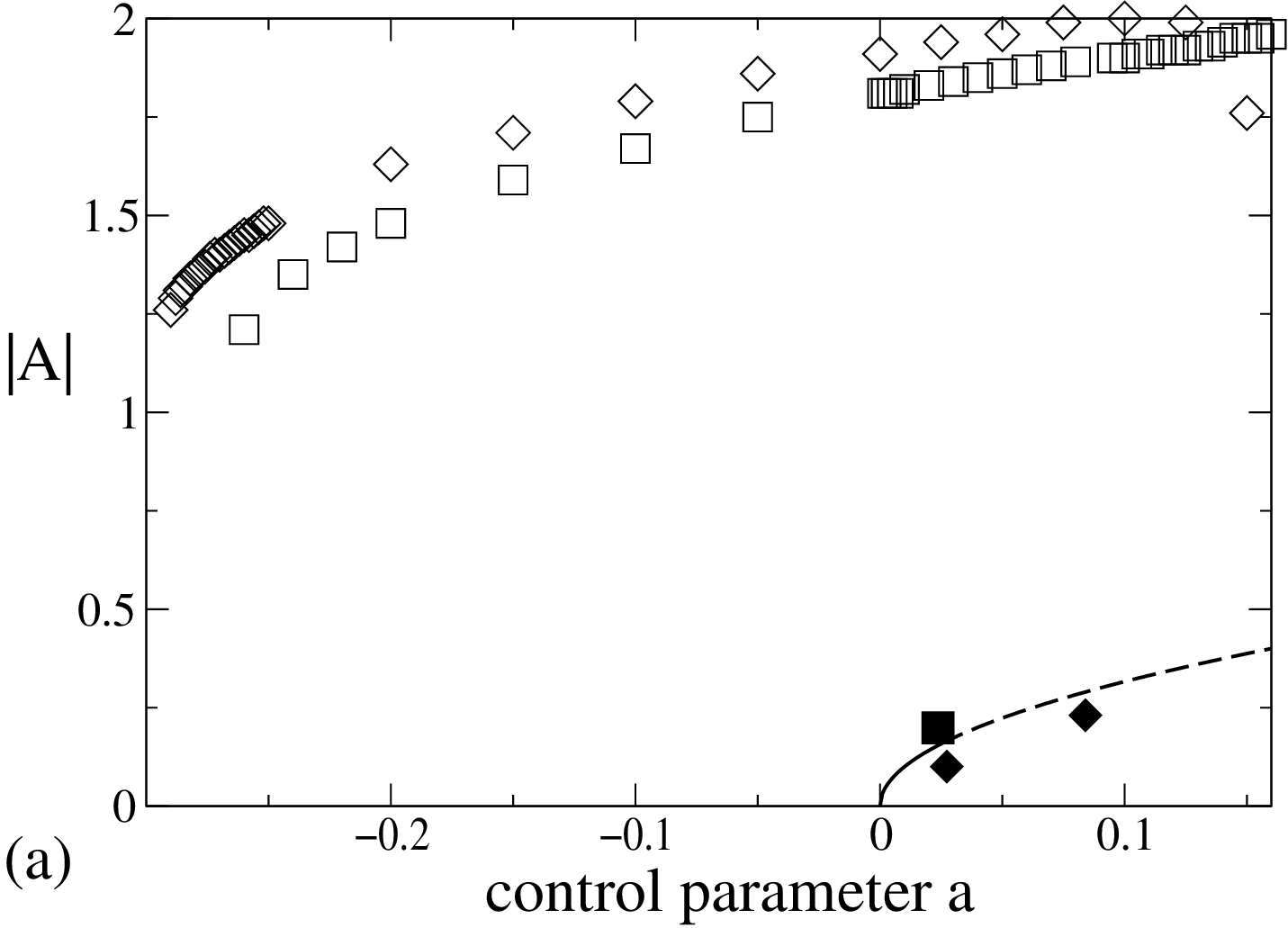}}\hspace{0.35in}\resizebox{3.2in}{!}{\includegraphics{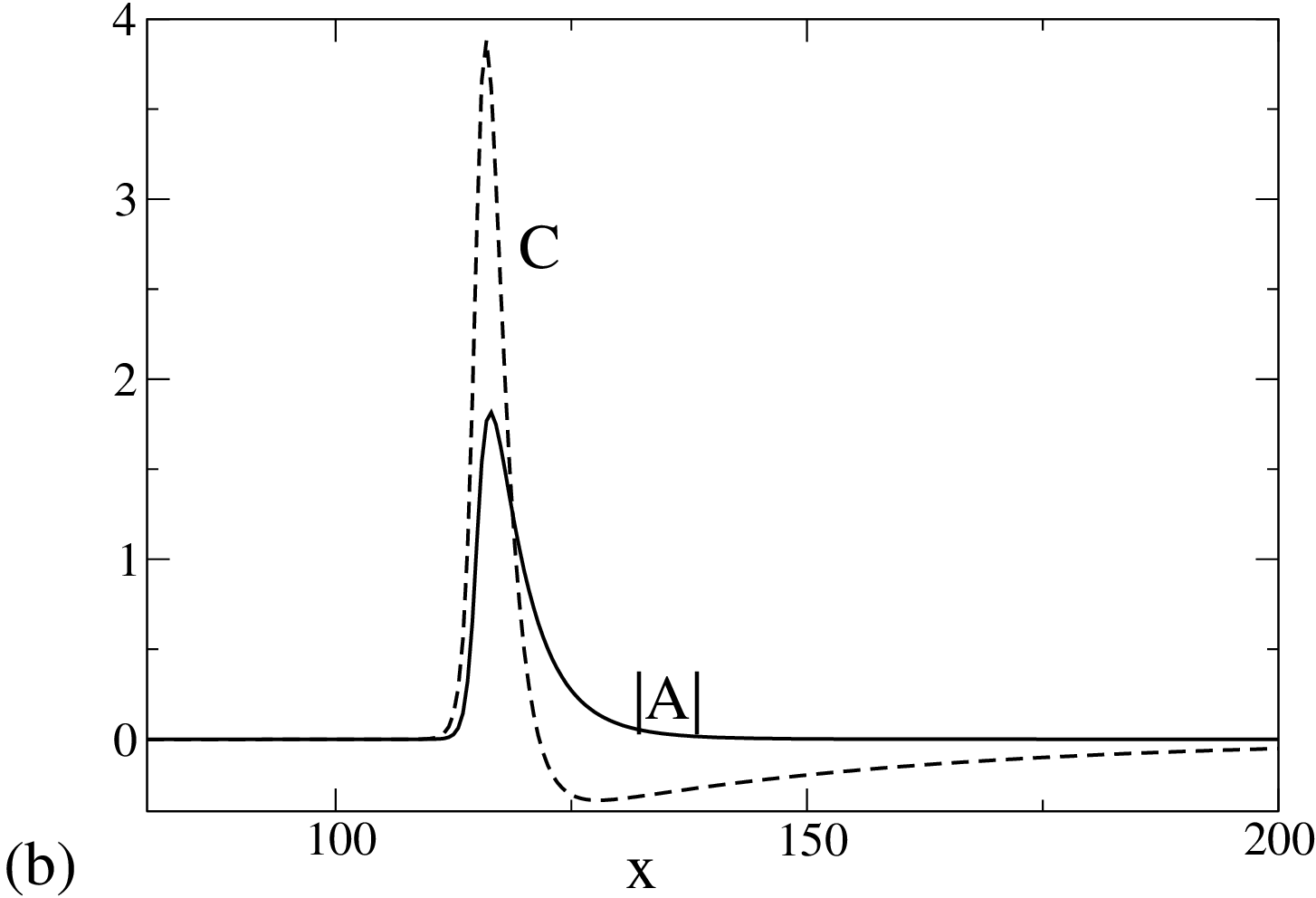}}}
\caption{(a) Bifurcation diagram of pulses; the subcritical branch extends over the conductive state.  The solid symbols indicate 
where the modes of the
corresponding branches become linearly unstable.
(b) Profile of the pulse solution with h=1, s=1, $\alpha$ = 0.02, $\delta$ = 0.5, a=-0.1. The solid line is the amplitude $|A|$ and 
the dashed line the C-field profile.} \label{Fig:bifpulse}
\end{figure}

\begin{figure}[t!]
\centerline{\resizebox{3.5in}{!}{\includegraphics{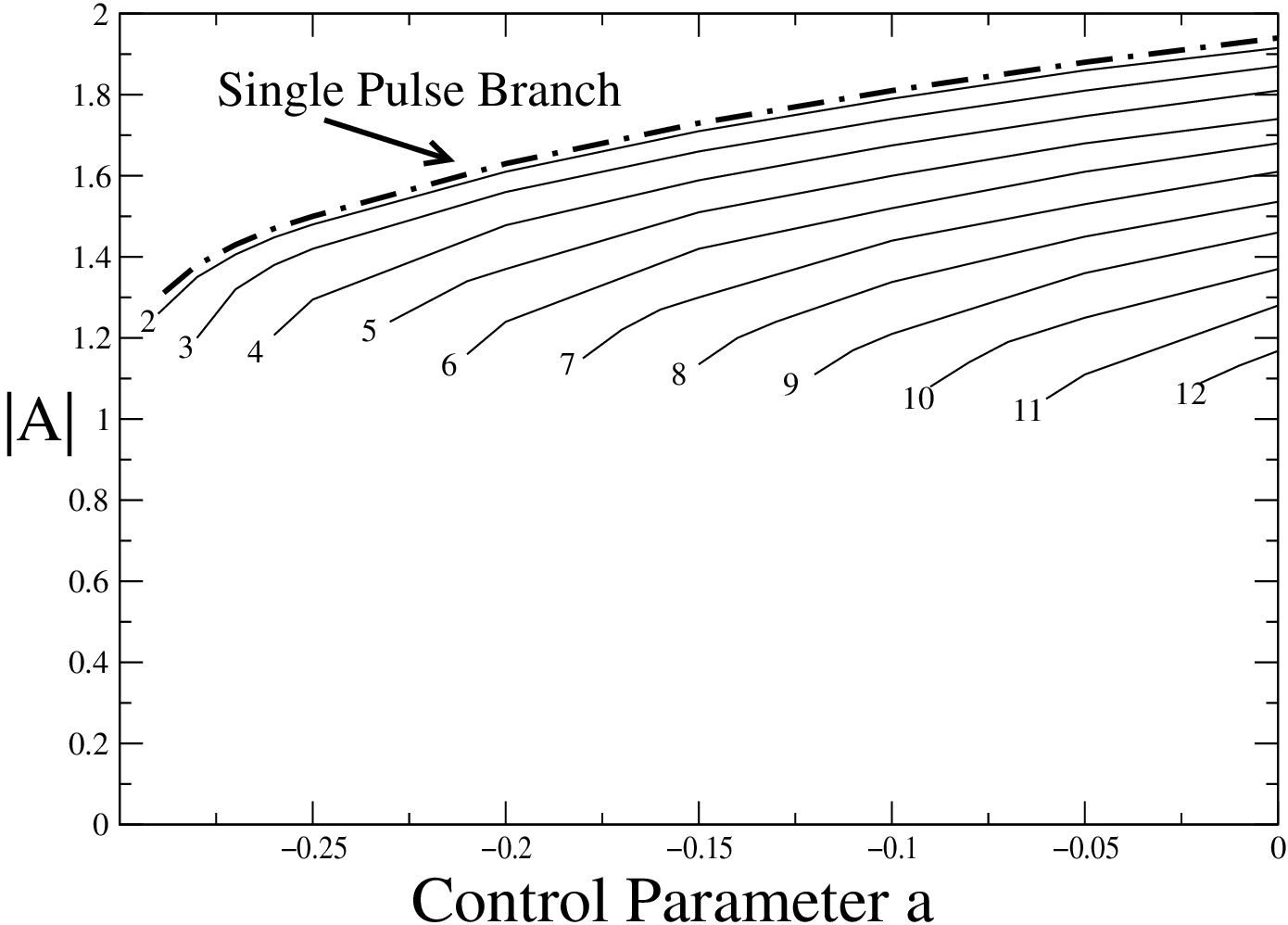}}}
\caption{Bifurcation Diagram with $h$=1.0, $s$=1.0, $\alpha$=0.02, $\delta$=0.5, $q$=0 and L=250.  Each branch represents a modulated-wave
solution with the number of modulations given.  As this number $n$ decreases, the modulated-waves become more like localized traveling
pulses.} \label{Fig:largesystem}
\end{figure}

As $\delta$ is decreased further, the subcritical branch of modulated waves, created in a saddle-node bifurcation as the control 
parameter is increased, extends further towards the onset of the traveling waves, eventually becoming bistable with the basic, 
conductive state (Fig. \ref{Fig:bifpulse}(a)).  As in the supercritical case, the modulated-wave branches arise from an instability of the plane-waves
to finite wavenumber modulations.  Modulation wavenumbers $p$ which lie in the
interval $p_{l}<p<p_{u}$, where the bounds depend on the value of the control parameter $a$, will linearly destabilize
the plane-waves.  For a finite system of length $L$, there will exist modulated-wave branches for all n such that $p=\frac{2n\pi}{L}$ 
lies in this interval.  As $n$ is decreased the solutions increasingly take on
the characteristics of localized objects (Fig. \ref{Fig:bifpulse}(b)).  In fact, the distance between them can become arbitrarily large.  Consequently, this
distance can be much larger than the maximal wavelength $\frac{2\pi}{p_{l}}$ of modulations that linearly destabilize the traveling waves in a given system.
Thus, these solutions do not bifurcate directly from the plane-wave branch but rather most likely arise in a tertiary bifurcation off a 
modulated-wave solution.  For example, Figure \ref{Fig:largesystem} shows a series of modulated-wave branches for a system (L=250) where a single modulation
 (n=1) does not linearly destabilize the waves.  Still, the single-pulse solution exists (dash-dot branch).

\section{Conclusion}

Systems similar to (3),(4) have been derived in various contexts, including a model of traveling interfacial waves in two-layer 
Couette flow \cite{ReRe93} where $\alpha=0$ due to conservation of the additional zero-mode, given by the position of
the interface.  This system was shown to exhibit a phase-instability leading to a phase-slip \cite{BaCh98} in agreement with
experiment \cite{ChBa99}.  A small-amplitude model for traveling waves in thermal, binary-mixture convection  consists
of complex Ginzburg-Landau equations for counterpropagating waves coupled to a slowly decaying mode representing large-scale 
variations
of the concentration field \cite{Ri92a}.  In this system the traveling waves arise in a backward bifurcation.  
Localized pulse solutions arising from instabilities of this system were
identified and many of their properties characterized \cite{Ri96}. 

 In this paper we have shown that the advected mode can cause the phase-instability to occur at finite wavelength
 and can introduce an additional amplitude-instability. The latter is identified as the origin of pulse-solutions seen in earlier
works.  They arise when the secondary bifurcation to modulated waves is sufficiently subcritical to lead to bistability between
the modulated wave and the basic state.
  A similar bifurcation may explain the appearance of localized ``worms'' in 
electroconvection in nematic liquid crystals, seen again before the primary instability to supercritical traveling waves.  Indeed,
a model of traveling waves coupled to a slowly decaying field has been invoked to describe the worm dynamics, attaining 
qualitative agreement \cite{RiGr98}.  

Finally, it would be worthwhile to consider the effect of dispersion in (\ref{EGLE:A}) on the dynamics discussed in this paper.  
In some cases dispersion has been shown to have a significant effect on the linear-stability properties of waves coupled to a additional mode.
  A model similar to (\ref{EGLE:A}, \ref{EGLE:C}) used by Bartelet and Charru \cite{ChBa99}, with $\alpha =0$ and added
dispersion, seems to explain qualitatively the linear-stability properties of waves observed in experiment.  The inclusion of 
dispersion also allows for a determination of the effect of the additional mode on the Benjamin-Feir instability.  In some cases
this effect is important as, for example, in the 4-species Oregonator model \cite{IpSo00} of the BZ reaction where the Benjamin-
Feir instability can be preceded by a finite-wavenumber instability.

We would like to thank Blas Echebarria for many interesting and helpful discussions. 
 This work was supported by the  Engineering Research Program of the
Office of Basic Energy Sciences at the Department of Energy
(DE-FG02-92ER14303) and by a grant from NSF ((DMS 9804673).

\section{Appendix A: Derivation of Amplitude Equation for Phase Instability}

Here we outline the derivation of the amplitude equation (\ref{LW:ampeq}) which describes the slow-time evolution of modulations
of plane waves in (\ref{LW:phi}, \ref{LW:C}).  We consider finite-wavenumber modulations $p_{cr}$ of
waves with wavenumber $q_{cr}$ as
given by (\ref{pcr}) and (\ref{qcr}).  We first expand the phase and C-field as,
\begin{equation}
\left( \begin{array}{cc}
\phi \\ C \end{array} \right)=
\left( \begin{array}{cc}
q_{cr}X \\ 0 \end{array} \right)+\epsilon\mathbf{\Phi_{0}}+\epsilon^{2}\mathbf{\Phi_{1}}+\dots,
\end{equation}
where $\epsilon =\sqrt{\frac{a-a_{0}}{a_{2}}}$ and $a_{0}$ is the value of the control parameter at onset of the fastest-growing mode. 
We rewrite  (\ref{LW:phi}, \ref{LW:C}) symbolically as,
\begin{equation}
\mathbf{\mathcal{L}}\left( \begin{array}{cc}
\phi \\ C \end{array} \right)+\mathbf{\mathcal{N}}(\phi ,C)=0,
\end{equation}
and expand the linear and nonlinear operators in $\epsilon$,
\begin{eqnarray}
\mathbf{\mathcal{L}}&=&L_{0}+\epsilon L_{1}+\epsilon^{2}L_{2}+\dots ,\\
\mathbf{\mathcal{N}}&=&\epsilon^{2} N_{1}+\epsilon^{3}N_{2}+\dots ,\\
D_{E}&=&D+\epsilon^{2}D_{2},
\end{eqnarray}
with
\begin{eqnarray}
L_{0}&=&\left( \begin{array}{cc}
D\partial_{X}^{2}-\partial_{T} & \frac{q_{cr}}{R^{2}}\partial_{X}\\
-2hq_{cr}\partial_{X}^{2} & \delta\partial_{X}^{2}-\partial_{T}-\alpha_{2}+(s+h)\partial_{X}\end{array}
\right),\\
L_{1}&=&0,\\
L_{2}&=&\left( \begin{array}{cc}
D_{2}\partial_{X}^{2}-\partial_{\tau} & R_{2}\partial_{X}\\
0 & -\partial_{\tau}\end{array}\right),\\
N_{1}&=&\left( \begin{array}{cc}
K_{1}C\partial_{X}^{2}\phi+K_{2}\partial_{X}\phi\partial_{X}^{2}\phi+K_{3}C\partial_{X}C+K_{4}\partial_{X}\phi\partial_{X}C,
& -2h\partial_{X}\phi\partial_{X}^{2}\phi\end{array}\right)^{T},\\
N_{2}&=&\left( \begin{array}{cc}
K_{5}\partial_{X}^{2}\phi (\partial_{X}\phi)^{2}+K_{6}C^{2}\partial_{X}^{2}\phi+K_{7}C\partial_{X}\phi\partial_{X}^{2}\phi
+K_{8}(\partial_{X}\phi )^{2}\partial_{X}C+\nonumber\\
K_{9}C\partial_{X}C\partial_{X}\phi
+K_{10}C^{2}\partial_{X}C, & 0\end{array}
\right)^{T}.\nonumber\\
\end{eqnarray}

where the coefficients are given by,
\begin{eqnarray}
R&=&\sqrt{a_{0}-q_{cr}^{2}},
R_{2}=-\frac{a_{2}q_{cr}}{R^{4}},\\
D&=&\frac{R^{2}-2q_{cr}^{2}}{R^{2}},
D_{2}=\frac{a_{2}}{R^{2}}(1-D),\\
K_{1}&=&\frac{(1-D)}{R^{2}},
K_{2}=\frac{2q_{cr}}{R^{2}}(D-3),\\
K_{3}&=&-\frac{q_{cr}}{R^{4}},
K_{4}=\frac{1}{R^{2}}(1+\frac{2q_{cr}^{2}}{R^{2}}),\\
K_{5}&=&\frac{1}{R^{2}}(D-3-\frac{8a_{0}q_{cr}^{2}}{R^{4}}),
K_{6}=-\frac{2q_{cr}}{R^{6}},\\
K_{7}&=&\frac{4q_{cr}}{R^{6}}(q_{cr}^{2}+a_{0}),
K_{8}=\frac{q_{cr}}{R^{4}}(3+\frac{4q_{cr}^{2}}{R^{2}}),\\
K_{9}&=&-\frac{1}{R^{4}}(1+\frac{4q_{cr}^{2}}{R^{2}}),
K_{10}=\frac{q_{cr}}{R^{6}}.
\end{eqnarray}

The slow time is given by $\tau =\epsilon^{2}T$.
At order $\epsilon$ we recover the homogeneous, linearized system,
\begin{equation}
L_{0}\mathbf{\Phi_{0}}=0. \label{NLphase1}
\end{equation}

The zero-mode $\mathbf{\Phi_{0}}$ we take, as in 3.1, to be of the form,
\begin{equation}
\mathbf{\Phi_{0}}=
\left( \begin{array}{cc}
\Phi_{0} \\ C_{0} \end{array} \right) A_{0}(\tau )e^{i(p_{cr}X+\omega_{cr}T)}+c.c.
\end{equation}

The eigenvector $(\Phi_{0}, C_{0})^{T}$ is determined by solving (\ref{NLphase1}),
\begin{equation}
\left( \begin{array}{cc}
\Phi_{0} \\ C_{0} \end{array}\right)=
\left( \begin{array}{cc}
iq_{cr}\frac{p_{cr}}{R^{2}} \\ Dp_{cr}^{2}+i\omega_{cr} \end{array}\right)
\end{equation}

The left-null eigenvector is found by solving the adjoint problem,
\begin{equation}
\mathbf{\Phi_{0}^{\dagger}}=
\left( \begin{array}{cc}
2hq_{cr}p_{cr}^{2} \\ Dp_{cr}^{2}-i\omega_{cr} \end{array}\right)e^{i(p_{cr}X+\omega_{cr}T)}.
\end{equation}

At O($\epsilon^{2}$) we have

\begin{equation}
L_{0}\mathbf{\Phi_{1}}+N_{1}(\mathbf{\Phi_{0}})=0. \label{NLphase2}
\end{equation}

The nonlinear term $N_{1}$ contributes both spatio-temporally homogeneous modes and modes of twice the 
critical modulation.  We thus take $\mathbf{\Phi_{1}}$ to be of the form,
\begin{equation}
\mathbf{\Phi_{1}}=
\left( \begin{array}{cc}
\Phi_{12} \\ C_{12} \end{array}\right)
A_{0}^{2}e^{2i(p_{cr}X+\omega_{cr}T)}+c.c+
\left( \begin{array}{cc}
T\Phi_{10} \\ C_{10} \end{array}\right)
|A_{0}|^{2}.  \label{NLphase2a}
\end{equation}

With this ansatz, (\ref{NLphase2}) reduces to two linear, algebraic systems.  The system corresponding to the forced second harmonic is
\begin{eqnarray}
\left( \begin{array}{cc}
-4Dp_{cr}^{2}-2i\omega_{cr} & 2ip_{cr}\frac{q_{cr}}{R^{2}}\\
8hq_{cr}p_{cr}^{2} & -4\delta p_{cr}^{2}-\alpha +2i(s+h)-2i\omega_{cr} \end{array}\right)
\left( \begin{array}{cc}
\Phi_{12} \\ C_{12} \end{array}\right)\nonumber\\
+\left( \begin{array}{cc}
K_{1}\Phi_{0}C_{0}p_{cr}^{2}+iK_{2}\Phi_{0}^{2}p_{cr}^{3}-iK_{3}C_{0}^{2}p_{cr}+K_{4}\Phi_{0}C_{0}p_{cr}^{2} \\ -2ih\Phi_{0}^{2}p_{cr}^{3}
\end{array}\right)=0,
\end{eqnarray}
and that for the homogeneous forcing
\begin{eqnarray}
\left( \begin{array}{cc}
-1 & 0\\
0 & -\alpha \end{array}\right)
\left( \begin{array}{cc}
\Phi_{10} \\ C_{10} \end{array}\right)+
\left( \begin{array}{cc}
-K_{1}p_{cr}^{2}(\Phi_{0}^{*}C_{0}+\Phi_{0}C_{0}^{*})+K_{4}p_{cr}^{2}(\Phi_{0}^{*}C_{0}+\Phi_{0}C_{0}^{*}) \\ 0
\end{array}\right)=0.
\end{eqnarray}
Note that the homogeneous forcing causes a response in the phase which grows linearly with time, i.e. it changes the frequency of
the wave (cf. (\ref{NLphase2a})).

At O($\epsilon^{3}$) we have

\begin{equation}
L_{0}\mathbf{\Phi_{2}}+L_{2}\mathbf{\Phi_{0}}+\mathbf{N_{1}}(\mathbf{\Phi_{0},\Phi_{1}})
+\mathbf{N_{1}}(\mathbf{\Phi_{1},\Phi_{0}})+\mathbf{N_{2}}(\mathbf{\Phi_{0},\Phi_{0}, \Phi_{0} })=0.
\end{equation}

At this order the critical Fourier mode appears through the nonlinear forcing terms and the 
action of the linear operator $L_{2}$ on the eigenvalue of the linearized operator of the
original system.  In order for a solution to exist, these contributions must lie orthogonal
to the left nullspace of the operator $L_{0}$.  We thus obtain the solvability condition,
\begin{equation}
\langle\mathbf{\Phi_{0}^{\dagger}}^{*}, L_{2}\mathbf{\Phi_{0}}+\mathbf{N_{1}}(\mathbf{\Phi_{0},\Phi_{1}})
+\mathbf{N_{1}}(\mathbf{\Phi_{1},\Phi_{0}})+\mathbf{N_{2}}(\mathbf{\Phi_{0}})\rangle =0, \label{NLphase3}
\end{equation}
where the inner product is defined as an integral over one period of the critical Fourier mode and the scalar product
of the scalar vectors in $\Phi, C$-space.  The solution of (\ref{NLphase3}) results in a differential equation (\ref{LW:ampeq})
for the complex amplitude $A_{0}$, the expression for which is extremely long and involved.

\section{Appendix B: Derivation of Amplitude Equation at Bandcenter for Amplitude-Driven Instability}

Here we follow the same procedure as in Appendix A in outlining the derivation of the Landau equation for the amplitude-driven 
instability.  The calculation is carried out at bandcenter, i.e. $q=0$.  We perturb the basic, plane-wave state in 
equations (\ref{BC:R}, \ref{BC:C}) by expanding the real amplitude R and real field C in 
$\epsilon =\sqrt{\frac{a-a_{0}}{a_{2}}}$ where $a_{0}$ is defined as in Appendix A,
\begin{equation}
\left( \begin{array}{cc}
R \\ C \end{array} \right)=
\left( \begin{array}{cc}
\sqrt{a_{0}} \\ 0 \end{array} \right)+\epsilon\mathbf{\Psi_{0}}+\epsilon^{2}\mathbf{\Psi_{1}}+\dots.
\end{equation}

We rewrite  (\ref{BC:R}, \ref{BC:C}) symbolically as,
\begin{equation}
\mathbf{\mathcal{L}}\left( \begin{array}{cc}
R \\ C \end{array} \right)+\mathbf{\mathcal{N}}(R ,C)=0.
\end{equation}
We now expand the linear and nonlinear operators in $\epsilon$,
\begin{eqnarray}
\mathbf{\mathcal{L}}&=&L_{0}+\epsilon L_{1}+\epsilon^{2}L_{2}+\dots ,\\
\mathbf{\mathcal{N}}&=&\epsilon^{2} N_{1}+\epsilon^{3}N_{2}+\dots ,
\end{eqnarray}
with
\begin{eqnarray}
L_{0}&=&\left( \begin{array}{cc}
\partial_{x}^{2}-\partial_{t}-2a_{0} & \sqrt{a_{0}}\\
2h\sqrt{a_{0}}\partial_{x} & \delta\partial_{x}^{2}-\partial_{t}-\alpha +s\partial_{x}\end{array}
\right),\\
L_{1}&=&0,\\
L_{2}&=&\left( \begin{array}{cc}
-2a_{2}-\partial_{\tau} & \frac{a_{2}}{2\sqrt{a_{0}}}\\
h\frac{a_{2}}{\sqrt{a_{0}}}\partial_{x} & -\partial_{\tau}\end{array}\right),\\
N_{1}&=&\left( \begin{array}{cc}
RC-3\sqrt{a_{0}}R^{2}, & 2hR\partial_{x}R\end{array}\right)^{T},\\
N_{2}&=&\left( \begin{array}{cc}
-R^{3}, & 0\end{array}
\right)^{T}.\nonumber\\
\end{eqnarray}

At order $\epsilon$ we recover the homogeneous, linearized system,
\begin{equation}
L_{0}\mathbf{\Psi_{0}}=0. \label{NLBC1}
\end{equation}

The zero-mode $\mathbf{\Psi_{0}}$ we take, as in 4.1, to be of the form,
\begin{equation}
\mathbf{\Psi_{0}}=
\left( \begin{array}{cc}
R_{0} \\ C_{0} \end{array} \right) A_{0}(\tau )e^{i(p_{cr}X+\omega_{cr}t)}+c.c.
\end{equation}

The eigenvector $(R_{0}, C_{0})^{T}$ is determined by solving (\ref{NLBC1}),
\begin{equation}
\left( \begin{array}{cc}
R_{0} \\ C_{0} \end{array}\right)=
\left( \begin{array}{cc}
\sqrt{a_{0}} \\ p_{cr}^{2}+i\omega_{cr}+2a_{0} \end{array}\right).
\end{equation}

The left-null eigenvector is found by solving the adjoint problem,
\begin{equation}
\mathbf{\Psi_{0}^{\dagger}}=
\left( \begin{array}{cc}
-2h\sqrt{a_{0}}ip_{cr} \\ p_{cr}^{2}+2a_{0}-i\omega_{cr} \end{array}\right)e^{i(p_{cr}X+\omega_{cr}T)}.
\end{equation}

At O($\epsilon^{2}$) we have

\begin{equation}
L_{0}\mathbf{\Psi_{1}}+N_{1}(\mathbf{\Psi_{0}})=0. \label{NLBC2}
\end{equation}

The nonlinear term $N_{1}$ contributes both spatio-temporally homogeneous modes and modes of twice the 
critical modulation.  We thus take $\mathbf{\Psi_{1}}$ to be of the form
\begin{equation}
\mathbf{\Psi_{1}}=
\left( \begin{array}{cc}
R_{12} \\ C_{12} \end{array}\right)
A_{0}^{2}e^{2i(p_{cr}X+\omega_{cr}T)}+c.c+
\left( \begin{array}{cc}
R_{10} \\ C_{10} \end{array}\right)
|A_{0}|^{2}.
\end{equation}

With this ansatz, (\ref{NLBC2}) reduces to two linear, algebraic systems.  The system corresponding to the forced second harmonic is
\begin{eqnarray}
\left( \begin{array}{cc}
-4p_{cr}^{2}-2i\omega_{cr}-2a_{0} & \sqrt{a_{0}}\\
4h\sqrt{a_{0}}ip_{cr} & -4\delta p_{cr}^{2}-2i\omega_{cr}-\alpha+2isp_{cr} \end{array}\right)
\left( \begin{array}{cc}
R_{12} \\ C_{12} \end{array}\right)\nonumber\\
+\left( \begin{array}{cc}
R_{0}C_{0}-3\sqrt{a_{0}}R_{0}^{2} \\ 2ihp_{cr}R_{0}^{2}
\end{array}\right)=0,
\end{eqnarray}
and that for the homogeneous forcing
\begin{eqnarray}
\left( \begin{array}{cc}
-2a_{0} & \sqrt{a_{0}}\\
0 & -\alpha \end{array}\right)
\left( \begin{array}{cc}
R_{10} \\ C_{10} \end{array}\right)+
\left( \begin{array}{cc}
R_{0}C_{0}^{*}+R_{0}^{*}C_{0}-6\sqrt{a_{0}}|R_{0}|^{2} \\ 0
\end{array}\right)=0.
\end{eqnarray}

Solving these systems yields the following expression for the scalar quantities,
\begin{eqnarray}
R_{12}&=&\frac{m_{12}}{d_{12}},\\
C_{12}&=&\frac{n_{12}}{d_{12}},\\
m_{12}&=&-\frac{1}{2}\sqrt{a_{0}} (4a_{0}\delta p_{cr}^{2}+4\delta p_{cr}^{4}+a_{0}\alpha
+p_{cr}^{2}\alpha +2\omega_{cr}sp_{cr}-2\omega_{cr}^{2}+i\lbrack 4\delta p_{cr}^{2}\omega_{cr}\nonumber\\
&&+\alpha\omega_{cr}-2a_{0}sp_{cr}-2sp_{cr}^{3}+2a_{0}\omega_{cr}+2\omega_{cr}p_{cr}^{2}+2a_{0}hp_{cr}\rbrack ), \\
n_{12}&=&-4a_{0}hp_{cr}\left( -\omega_{cr}+i\lbrack a_{0}+\frac{3}{2}p_{cr}^{2} \rbrack \right), \\
d_{12}&=&-8\delta p_{cr}^{4}-2(\alpha +2a_{0}\delta )p_{cr}^{2}-a_{0}\alpha +2\omega_{cr}(\omega_{cr}-sp_{cr})\nonumber\\
&&i\lbrack 4sp_{cr}^{3}-4\omega_{cr}(1+\delta )p_{cr}^{2}+2a_{0}p_{cr}(s+h)-\omega_{cr}(2a_{0}+\alpha)\rbrack ,\\
R_{10}&=&-\sqrt{a_{0}}+\frac{1}{\sqrt{a_{0}}}(p_{cr}^{2}+2a_{0}),\\
C_{10}&=&0.
\end{eqnarray}

At cubic order projection onto the left null eigenvector yields the amplitude equation for $A_{0}$,
\begin{eqnarray}
\partial_{\tau}A_{0}&=&l(p_{cr})(p_{cr}^{2}+i\omega_{cr})A_{0}+l(p_{cr})\sqrt{a_{0}}\nonumber\\
&&\lbrack C_{12}\sqrt{a_{0}}
-3a_{0}^{3/2}+2R_{12}(p_{cr}^{2}-a_{0})+2R_{10}(p_{cr}^{2}-a_{0}+i\omega_{cr})\rbrack |A_{0}|^{2}A_{0},\\
l(p_{cr})&=&2ihp_{cr}\frac{\lbrace (p_{cr}^{2}+2a_{0})^{2}-\omega_{cr}^{2}-2i\lbrack \omega_{cr}(p_{cr}^{2}+2a_{0})
+ha_{0}p_{cr}\rbrack\rbrace}{\lbrack (p_{cr}^{2}+2a_{0})^{2}-\omega_{cr}^{2}\rbrack^{2}+4\lbrack \omega_{cr}
(p_{cr}^{2}+2a_{0})+ha_{0}p_{cr}\rbrack^{2}}.
\end{eqnarray}

\bibliography{/stash/alex/.bibfiles/journal}
\end{document}